# Display Content, Display Methods and Evaluation Methods of the HCI in Explainable Recommender Systems: A Survey

Weiqing Li, Yue Xu, Yuefeng Li, and Yinghui Huang*

**Abstract:** Explainable Recommender Systems (XRS) aim to provide users with understandable reasons for the recommendations generated by these systems, representing a crucial research direction in artificial intelligence (AI). Recent research has increasingly focused on the algorithms, display, and evaluation methodologies of XRS. While current research and reviews primarily emphasize the algorithmic aspects, with fewer studies addressing the Human-Computer Interaction (HCI) layer of XRS. Additionally, existing reviews lack a unified taxonomy for XRS and there is insufficient attention given to the emerging area of short video recommendations. In this study, we synthesize existing literature and surveys on XRS, presenting a unified framework for its research and development. The main contributions are as follows: 1) We adopt a lifecycle perspective to systematically summarize the technologies and methods used in XRS, addressing challenges posed by the diversity and complexity of algorithmic models and explanation techniques. 2) For the first time, we highlight the application of multimedia, particularly video-based explanations, along with its potential, technical pathways, and challenges in XRS. 3) We provide a structured overview of evaluation methods from both qualitative and quantitative dimensions. These findings provide valuable insights for the systematic design, progress, and testing of XRS.

## Acknowledgment

This research is supported by Program of National Natural Science Foundation of China (Grant no: 72304090 and 72204095). And we would like to sincerely thank the reviewers for their valuable insights, which greatly enriched the content of this paper and enhanced its depth and significance.

---

- Weiqing Li is with School of Economics and Management & Hubei Research Center for Digital Industrial Economy Development, Hubei University of Technology, Wuhan, 430079, China.
- Yue Xu and Yuefeng Li are with Faculty of Science, School of Computer Science, Queensland University of Technology, Brisbane, 4000, Australia.
- Yinghui Huang is with School of Management, Wuhan University of Technology, Wuhan, 430079, China. E-mail: huangyh@whut.edu.cn
- * To whom correspondence should be addressed.

# 1    Introduction

Recommendation systems are becoming increasingly popular and important in many domains, such as e-commerce, social media, and healthcare. These systems use algorithms to make personalized recommendations to users, which can significantly impact user behavior and decision-making. However, users often have little understanding of how these recommendations are generated, leading to mistrust and frustration. The explainable recommender systems (XRS) aim to address this issue by presenting the reasons for recommending or not recommending an item, or providing users with explanations of how recommendations are generated (i.e., the logic of underlying algorithms), which can be presented in the form of text, pictures, animation, video, etc [1, 2]. There is a growing demand for explainable artificial intelligence (XAI), where systems can provide reasons behind their recommendations. The explainability of recommendation systems helps researchers and developers understand algorithms, debug models, and control results. It can increase user trust, improve user experience, and help users make better decisions [3, 4].

**What is Human-Computer Interaction (HCI) in XRS?** HCI is the interdisciplinary field that studies the design, evaluation, and implementation of interactive computing systems for human use and the major phenomena surrounding them [5]. HCI focuses on how people interact with computers and other technology systems, aiming to improve the usability, efficiency, and user experience of these interactions. It combines elements of computer science, psychology, design, and ergonomics to optimize the way people use and benefit from technology [6]. In the context of XRS, HCI focuses on how explanations of recommendations are presented to users, how users perceive and understand the recommendation, explanations, even algorithm, and how these interfaces can improve user trust, satisfaction, and decision-making [7].

**Why we care about the HCI in XRS ?** The XRS can encourage user engagement and interaction.

Certain researchers classified existing XRS research into two orthogonal dimensions: 1) the models or algorithms to generate explanations, which represents the machine learning (ML) perspective [8]. For example, collaborative-based XRS algorithms [9]; explicit factorization-based XRS algorithms [10]; multi-task learning XRS algorithms [11]; knowledge graph-based XRS algorithms [12]; and more recently, there have been proposals for causality-based XRS algorithms [13, 14]. 2) The information source or display style of the explanations, which represents the HCI perspective. HCI plays a key role in making complex XRS models more understandable. It serves as the bridge between the system's technical processes and the user's comprehension, ultimately influencing trust, satisfaction, and the success of the recommendations.

It is evident that research on models for XRS has garnered widespread attention and lots of models and algorithms for XRS have been proposed. However, existing research on the HCI aspect of XRS is somewhat lacking, primarily in the following areas:

**First, the taxonomies of XRS research are disorganized.** While there exists a wide variety of taxonomies based on different dimensions, they are neither well-standardized nor structured with a clear category hierarchy [8,15,16]. These taxonomies [8, 15-18] emphasize varying perspectives but lack the unified primary categories seen in recommendation systems, such as the widely accepted collaborative-based, content-based, and hybrid taxonomy. Furthermore, the current taxonomies lack cohesion and hierarchy, often causing confusion and difficulty for readers and practitioners in understanding and applying. This issue will be discussed in detail in Section 3.3.

**Second, there is insufficient attention to the application of cutting-edge technologies in XRS.** While the widespread and successful application of multimedia such as video and audio across various platforms [19, 20], there is noticeable neglect of research focusing on multimedia-based explainable recommendation methods, which have not been sufficiently covered in existing taxonomies.

Therefore, after summarizing and synthesizing the existing literature and reviews, as well as comparing and analyzing existing taxonomies in the field of XRS, this article provides a comprehensive review of the display content, methods, and evaluation of the HCI layer in XRS, contributing to the following aspects.

First, we propose a comprehensive framework that covers the entire process of XRS, systematically summarizing the technologies and methods used in XRS. This framework not only addresses the inconsistencies and limitations of existing XRS taxonomies but also provides a clearer structure for understanding the interplay between algorithms, explanation models, and user interactions.

Second, we emphasize the inclusion of multimedia, particularly video-based explanations, as a novel dimension in XRS research. Videos provide rich visual and auditory information that can intuitively convey complex relationships and product details, making them highly effective for improving user engagement and comprehension. We also explore the potential, technical paths, and challenges of video-based explanations, offering valuable insights for future advancements.

Third, we present a structured overview of evaluation methods for XRS from both qualitative and quantitative dimensions, highlighting key gaps such as the lack of standardized scales and methods to assess responsiveness and cognitive load, offering directions for advancing HCI theories and improving XRS evaluation.

The remainder of this paper is organized as follows: In Section 2, we introduce the review methodology used in this survey. In Section 3, we outline the background and analyze related work to propose a framework. We then conduct a comprehensive examination of existing HCI research for XRS from three perspectives. Section 4 focuses on display content for XRS, Section 5 addresses display methods for XRS, and Section 6 covers display evaluation for XRS. Finally, Section 7 summarizes this work and provides insights and guidelines for future research.

## 2 Review Methodology

In this section, we describe the process we followed in designing our survey, focusing on display content, methods, and evaluation of HCI in the context of XRS.

### 2.1 Research Goal

The goal of this review is to provide researchers and system designers with a global perspective to comprehensively understand the development and evaluation of HCI applications in XRS.

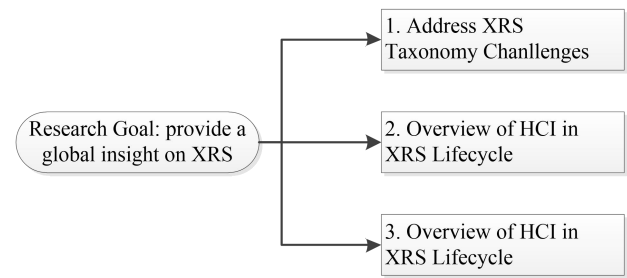

**Fig. 1 Research Goal**

As illustrated in Fig. 1, we aim to: 1) address the challenges in existing XRS taxonomies due to inconsistency and lack of clarity; 2) provide a detailed overview of the display content, methods, and evaluation approaches for HCI from a lifecycle perspective of XRS; and 3) supplement existing research with advanced approaches, specially, the video-based explanation methods, by exploring their potential, application scenarios, implementation pathways, and associated challenges in XRS.

### 2.2 Search Methodology

We followed the PRISMA methodology to search and select literature for this review (Fig. 2). We selected four main online publication databases pertinent to recommender system research: ACM Digital Library, IEEE Xplore Digital Library, ScienceDirect, and Springer Link. The search query used was (HCI OR display OR presentation OR visualization) AND (explanation OR explainable) AND (recommender system OR recommendation). 489 articles were selected from these 4 databases. Additionally, we searched other databases, such as Google Scholar and arXiv, and 9 articles were supplemented. After excluding duplicates, 272 papers remained.

In the third step, we applied four exclusion criteria (EC) and five inclusion criteria (IC) to select papers relevant to the primary studies for analysis in our review (Fig. 2). Papers were excluded if they were not written in English (EC1), lacked full-text access (EC2), were duplicates of more complete studies (EC3), or focused solely on algorithms without addressing explanation presentation (EC4). Conversely, we included studies that focused on explainable recommendation systems with an emphasis on explanation display (IC1), presented a software application featuring recommendation explanations or descriptions (IC2), or evaluated explanation methods (IC3). Additionally, to gain a more comprehensive understanding of XRS from a lifecycle perspective, we included classic papers on explainable recommendation algorithms and models (IC4) and systematic reviews (IC5). Ultimately, a total of 102 primary works were included in the final analysis.

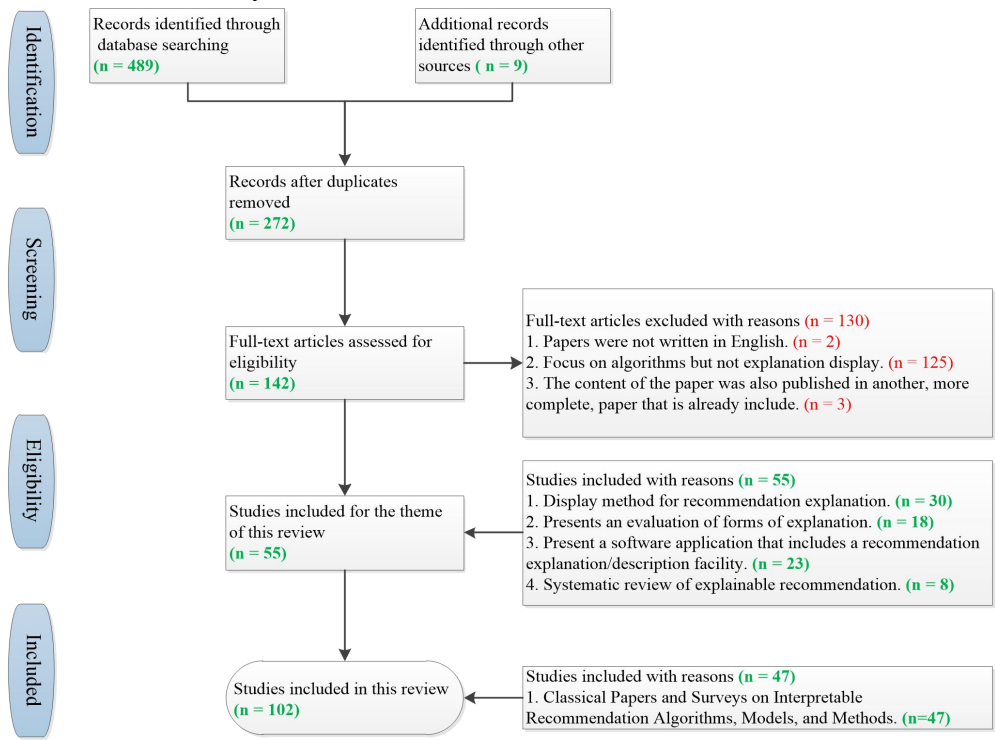

**Fig. 2 Search methodology of PRISMA**

# 3    Research Background

In this section, we will begin with a literature review of XRS research from a historical perspective, followed by an overview of existing explainable recommendation algorithms and models. Finally, we will conduct a comprehensive comparison and analysis of existing surveys and taxonomies.

## 3.1    Background

Before the concept of XRS was proposed, scholars suggested that recommendations could be explained through other items familiar to the user [21] or by referencing similar neighbor users of the target user [17], These approaches were applied in systems like Amazon's product recommendation system and are essentially the fundamental ideas behind item-based (IB) and user-based (UB) collaborative filtering (CF). Herlocker et al. later proposed using a rating histogram of neighboring users to explain the user-based collaborative filtering method in MovieLens, resulting in higher user satisfaction [9]. In 2007, Tintarev et al. provided a comprehensive review of explanations in recommender systems, highlighting seven potential advantages of explanation facilities. These seven criteria have been widely adopted, significantly advancing the development of XRS [22]. In the following years, efforts were made to enhance explanation interfaces, making them more detailed and comparable to the tags or features of recommended items. For instance, Pu and Chen proposed an 'Organization Interface' that not only provides the most popular product but also recommends 24 other candidates with a 'why' tooltip, allowing users to compare each candidate to the recommended product based on different features [23]. Some other scholars have also explored the standardized or graphical display methods for features, such as Tagsplanation, which introduces the relevance and preference of a movie's feature [24]. Word cloud explanations display a collection of keywords relevant to the recommended item [25]. Horizontal bar explanations [26] or radar chart explanations [27] quantify features and facilitate comparisons across candidates. Additionally, several researchers have suggested methods for generating explanatory sentences based on tags or features, including template-based explanations, which define sentence templates filled them with different words for different users [28, 29], and generating sentence explanations using natural language processing methods to create reasonable explanations [30, 31].

In recent years, more complex, diverse, and interactive interfaces have been proposed for XRS. These interfaces provide multidimensional explanations, varying levels of explanations, and even describe the principles of the recommendation algorithms to the user. Controllable and explainable interfaces allow users to adjust the weighting of recommendation models, enabling them to tailor the recommended items according to their needs and preferences. This can improve recommendation effectiveness and efficiency [32-38]. However, user-controlled interfaces for the hybrid recommender systems cannot guarantee that users understand the underlying rationale of the recommendation algorithm [39]. Consequently, these interfaces are often applied in more specialized areas, such as academic networks.

In addition, there are numerous applications that utilize videos to showcase recommendation lists, such as Netflix, Taobao, and Amazon. This approach combines visual and auditory elements to provide users with a more intuitive presentation of recommended content, enhancing their understanding and interest, thereby achieving the goal of explaining recommendations to users [40].

## 3.2 Algorithms of XRS

XRS can be implemented through two main approaches: model-based and post-hoc methods [8].

### 3.2.1 Model-based explanation methods

Model-based (also called ante-hoc or intrinsic) explanation methods derive explanations directly from the internal recommendation algorithm or learned parameters of a recommendation model. These explanations often leverage the mathematical or algorithmic foundations of the model to reveal why a specific recommendation was made [41]. Examples include:

**Demographic-based recommendation algorithms**: Among the earliest approaches, demographic-based recommendation algorithms leverage user characteristics—such as age, gender, occupation, education, and location—to generate recommendations [42, 43].

**Rule-based recommendation algorithms**: also known as Knowledge-based recommendation [44], are a class of recommendation algorithms that rely on predefined rules and heuristics to generate suggestions for users. Unlike data-driven models, these systems operate using explicit logic defined by domain experts or business requirements. They are often employed in scenarios where interpretability, simplicity, or adherence to specific business constraints is critical. Rule-based systems excel in environments with structured and well-understood data but may lack adaptability in dynamic, large-scale contexts [45].

**Content-Based Recommendations:** These methods generate recommendations by analyzing item characteristics to construct user preference profiles based on interaction data. Recommendations are derived by computing similarities between item and user profiles using models like TF-IDF, decision trees, clustering, or cosine similarity [46]. Unlike collaborative approaches, these methods are independent of other users' preferences. Explanations highlight similarities between recommended and previously liked items and the weight of recurring features representing user preferences [47, 48].

Then we will introduce collaborative recommendation algorithms and their interpretability, termed "collaborative" as they leverage user-derived information for generating recommendations and explanations. These algorithms are broadly categorized into two types: memory-based and model-based approaches. Memory-based methods include user-based and item-based collaborative filtering, while model-based approaches encompass association rule-based recommendations, latent matrix factorization, deep learning models, and knowledge graph-based models.

**User-Based Collaborative Filtering (UBCF):** Recommends items to a user based on the preferences of similar users. The similarity is typically calculated using measures like cosine similarity, Pearson correlation, or Jaccard similarity [49].

**Item-Based Collaborative Filtering (IBCF):** Recommends items by analyzing similarities between items based on user ratings. It focuses on the relationship between items rather than users [50].

**Association Rule-Based Recommendations:** Commonly referred to as market basket analysis, this approach identifies frequent itemsets in large datasets to infer associations [51]. The interpretability of this method stems from the high co-occurrence frequency of recommended items with previously purchased or liked items. For instance, 10% of shoe buyers also purchase socks, and 60% of bread buyers purchase milk, as illustrated by the well-known "diapers and beer" case [52, 53]. In this study, we classify association rule-based recommendation algorithms under model-based systems, diverging from prior research that categorized them as rule-based systems. This classification is supported by two key arguments: 1) Association rules are dynamic, generated through frequent pattern mining in large datasets, rather than predefined static rules; 2) Frequent pattern mining determines the co-occurrence probability of items based on the purchase histories of all users, aligning with the collaborative paradigm.

**Explicit Factor Model (EMF)-based**

**recommendation**: These models recommend products that excel in features aligned with user preferences by extracting explicit product features (e.g., screen, speaker) and user preferences from user behavior, item attributes, and reviews [10, 54]. By constructing user feature-preference pairs, the model provides a clear mapping between users and the features they value. Built on latent factor models (LFM), each latent dimension corresponds to an explicit feature, ensuring interpretability at both the algorithmic and feature levels [10, 55].

**Knowledge graph-based models:** Use graph embeddings to represent entities and their relationships, enriching semantic representations of users and items [56]. Explanations are generated by computing similarities between nodes [57] or identifying paths within the graph [58].

**Deep learning-based methods:** Often have complex hierarchical structures, making direct interpretation challenging [59]. Explanations are typically provided via attention mechanisms, which highlight the features contributing most to the recommendations [60]. Alternatively, post-hoc explanation methods can be employed to interpret results without redesigning or retraining the model [61, 62].

Additionally, recent advancements have introduced **causal inference-based models** for explainable recommendation. These models incorporate causal reasoning into recommendation systems to identify causal relationships rather than simple correlations. This approach aims to deepen the understanding of user behavior and recommendation mechanisms, thereby providing meaningful and interpretable explanations for both users and the system [63]. The key methodologies include 1) Causal Effect Analysis: Examines the impact of specific variables on outcomes to elucidate causal influences [64]; 2) Counterfactual Analysis: Explores hypothetical scenarios to explain how alternative conditions could affect recommendations [65]; 3) Debiasing Strategies: Mitigates biases in data to improve the validity and fairness of causal interpretations [66].

### 3.2.2 Post-hoc explanation methods

Post-hoc (also called model-agnostic) explanations provide interpretable insights into recommendation outcomes without modifying the underlying recommendation algorithms. These approaches operate as supplementary models, elucidating the rationale behind predictions after their generation [67]. Post-hoc explanation methods can be categorized into three primary types: feature-importance-based methods, content analytic-based methods, and natural language generation-based methods [68].

**Feature-Importance-Based Methods:** These methods identify and rank features influencing recommendation outcomes. Two prominent techniques include LIME (Local Interpretable Model-Agnostic Explanations) and SHAP (SHapley Additive explanations). LIME approximates complex models with interpretable surrogate models in a localized context. By perturbing input data and analyzing the resulting output variations, it identifies the most influential features (e.g., user demographics, item attributes) driving a specific recommendation. SHAP applies game-theoretic principles to compute Shapley values, which quantify the marginal contribution of each feature to a prediction. It offers consistent, additive explanations by breaking down recommendation scores into individual feature contributions [69].

**Content Analytic-Based Methods:** Justify recommendations by analyzing and summarizing the content of items or user profiles. These methods rely on textual or multimedia data (e.g., reviews, tags, descriptions) associated with items to uncover features or patterns aligned with user preferences [70]. Examples like feature-based explanations using word clouds [71].

**Natural Language Generation-Based Methods.** Natural language generation (NLG) techniques craft personalized explanations using identified features or patterns. These methods primarily include 1) Template-Based Methods: Predefined templates structure explanations, ensuring clarity and consistency [72]; 2) GPT-Based Methods which Advanced NLP models, such as GPT, generate dynamic, context-aware explanations tailored to individual recommendations [73].

In conclusion, XRS encompass a diverse range of algorithms and methods, from traditional to advanced, and from simple to complex. The field has evolved rapidly, demonstrating both the complexity and continuous innovation aimed at improving interpretability and enhancing user understanding.

### 3.3    Related Surveys and Taxonomies

Several researchers have conducted literature reviews on relevant studies and proposed taxonomies to classify explainable recommendations from various perspectives. In this section, we provide an overview of these studies and compare their taxonomies to identify similarities and differences.

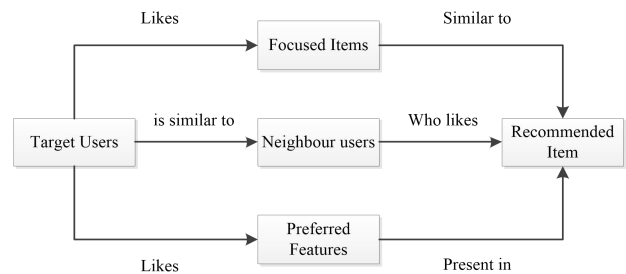

**Fig. 3 Intermediary entities (center) relate user to recommended item [24]**

Vig et al. [24] introduced a taxonomy where explanations in recommender systems rely on an intermediary entity to establish the connection between the user and the recommended item. This taxonomy classifies explanations into three types. As depicted in Fig. 3, **1) item-based:** The recommendation and explanation are justified by highlighting the similarity between the recommended item and the focused item that the target user has previously liked; **2) user-based:** the recommendation is explained by identifying a neighbor user who shares similar preferences with the target user and has liked the recommended item; **and 3) feature-based,** The explanation is derived from the presence of preferred features in the recommended item, aligning with the characteristics that the target user has shown interest in. Each type corresponds to the specific intermediary entity utilized to link the user to the recommended item. Expanding upon this work, Papadimitriou et al.

[18] proposed a more comprehensive taxonomy of explanation types (styles) in recommender systems, categorized into four groups: **1) Human style of explanation,** which provides explanations based on similar users; **2) Item style of explanation,** based on choices a user has made for similar items; **3) Feature style of explanation,** which draw on item features previously rated by the user; **4) Hybrid style of explanation,** combining elements from the aforementioned styles [18].

Friedrich and Zanker [17] developed a taxonomy that organizes explanation approaches across three dimensions: **reasoning model, recommendation paradigm, and exploited information.** The reasoning model pertains to how explanations are generated, while the recommendation paradigm includes collaborative filtering, content-based filtering, and knowledge-based recommendations. The exploited information dimension considers three primary types of input: the user model, which includes known preferences, ratings, or demographics; the recommended item, focusing on its specific characteristics; and alternatives, highlighting comparative advantages and disadvantages relative to the recommended item.

Nunes and Jannach [15] conducted a comprehensive review of explanations in decision-support systems, focusing on the processes for algorithmic generation, the types of information employed, and the methods of presentation. Their taxonomy identifies four key dimensions for designing explanations: **1) explanation objectives,** which encompass goals such as transparency, scrutability, trust, effectiveness, persuasiveness, efficiency, and satisfaction; **2) explanation responsiveness,** which reflects the level of detail tailored to user expertise or preferences; **3) explanation content,** which includes elements such as user preferences, inference processes, contextual or complementary information, and comparisons with alternatives; and **4) explanation presentation,** which refers to delivery formats, including natural language, visualizations, lists, logs, and arguments.

More recently, Mohamed et al. [16] classified visually XRS based on four dimensions: **1) Explanation goal**, align with the definition provided in reference [15]; **2) Explanation scope**, the explainability of the recommendation input (i.e., user model), the recommendation process (i.e., algorithm), and/or the recommendation output (i.e., product); **3) Explanation style,** focus on strategies for generating explanations, categorized into content-based, collaborative-based (further divided into neighborhood-based and model-based), social, and hybrid methods [74]; **4) Explanation format,** describe the presentation styles encompassing textual explanations and visual explanations. Recommendation explanations can be presented in very different display styles, which could be a relevant user or item, an image, a sentence, a chart, or a set of reasoning rules.

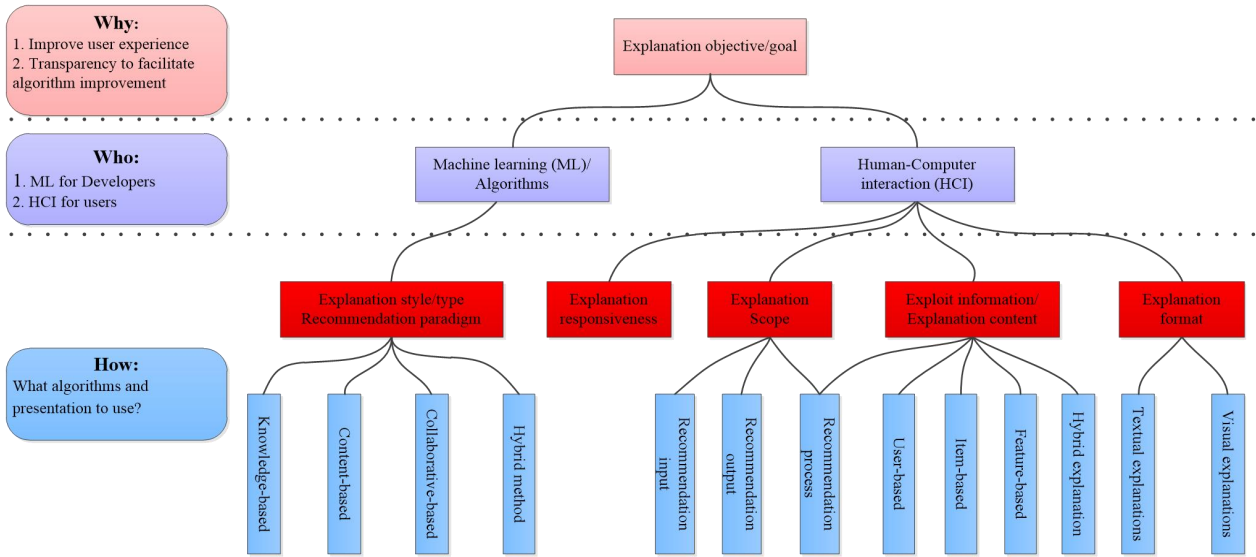

**Fig. 4 A summary of the taxonomies and categories in XRS**

Based on the taxonomies discussed in the aforementioned literature and aligned with the three foundational questions in XAI (Why, Who, and How) [75]. We attempt to hierarchically classify and unify the various categories, as illustrated in Fig. 4.

At the first level, the purpose of explanation (Why) is twofold: first, to enhance users' comprehension and trust in the system, thereby improving the overall user experience; second, to promote algorithmic transparency, which facilitates system debugging and development. Building on these objectives, the second level identifies the target audience of the explanations (Who): ML-oriented explanations primarily serve developers, whereas explanations grounded in HCI—are designed for end-users [76]. The third level addresses the implementation of the explanation (How), focusing on the techniques and methodologies employed to generate explanations. Aforementioned taxonomies primarily focused on this layer, as highlighted in red in Fig. 4.

However, we argue that these taxonomies present certain limitations and confusion. For example, **1) Terminological inconsistency:** terms such as "explanation style/type" [16, 18] and "recommendation paradigm" [17] are both used to denote algorithm-level design choices in XRS, yet lack a coherent and unified conceptualization; **2) Introduction of abstract concepts without corresponding implementation**

strategy — "explanation responsiveness" [15]; **3) Divergent classification perspectives applied to identical techniques**—such as the reclassification of five methods under "explanation content" into three tiers (input, output, and process) within the "explanation scope" framework [16]. Accordingly, current literature on XRS taxonomies introduces considerable conceptual complexities and ambiguities.

### 3.4 Summary and Commentary

In summary, it is imperative to consolidate and unify their underlying concepts before proceeding with further analysis. It is challenging to examine HCI aspects in isolation from the algorithms that drive them. Attempting to employ a single taxonomy or a limited set of taxonomies to capture all dimensions of XRS proves insufficient due to the diversity and complexity of existing XRS. For instance, systems like "Knowledge Graph-Based Causal Recommendation (KGCR) [77]" integrate knowledge graph techniques, causal modeling, and post-hoc explanation methods.

Therefore, we move beyond classification and instead adopt a process-oriented approach to clarify the techniques and methods employed in different XRS. Specifically, the content of explanations serves as input data that, when processed by designated algorithms, is subsequently presented to users. As the logical principles underpinning recommendation algorithms are integral to the HCI presentation

layer—forming the foundation for generating recommendation outputs. To address this gap, building on prior studies, established taxonomies, and the complete lifecycle of XRS development, we propose a four-stage framework for XRS advancement, as illustrated in Fig. 5: data input, recommendation algorithms, explanation display, and evaluation. At the conclusion of the study, we will revisit and refine this framework, incorporating insights from the review to present a finalized version in Fig. 29.

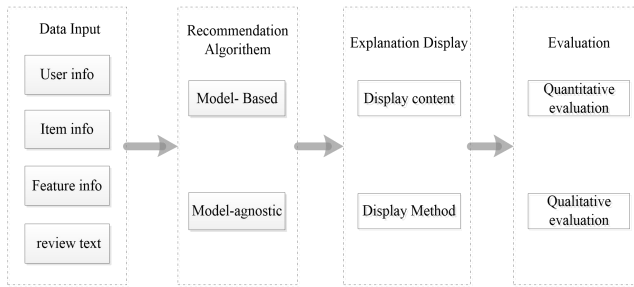

**Fig. 5 Four-stage framework for XRS**

# 4     Display Content of XRS — Data Input Processed by Algorithms

Display content for XRS refers to the core information presented after input data (user information, item information, feature information, review text) has been processed by the recommendation algorithm or explanation model. As shown in Fig. 4, display content can be categorized into five types: user-based, item-based, feature-based, recommendation process, and hybrid explanations.

To validate the rationale behind this taxonomy, we first analyze it from the perspective of "data input," focusing particularly on "review text" and the "recommendation process." In XRS research, the handling of "review text" often involves extracting positive user reviews about specific product features, which are then used to generate feature-level explanations. The term "recommendation process," on the other hand, refers to the description of how recommendations are generated, including the specific rules within rule-based recommendation models or the underlying principles of recommendation algorithms. Based on this interpretation, we propose that "logical-based explanation" is a more precise term for this category.

Thus, the final classification of display content for XRS is as follows: user-based, item-based, feature-based, logical-based, and hybrid explanations.

## 4.1     User-based Explanations

User-based recommendation explanations are predominantly derived from users' implicit or explicit feedback, such as demographic attributes, ratings, purchase history, search activities, or social connections. These recommendations and explanations are often generated using demographic-based or neighboring user-based methods.

Demographic-based recommendations often employ demographic triggers to construct explanations, such as suggesting that a specific product aligns with a user's professional needs, targeting cosmetic products toward women [78], or recommending winter apparel as the season transitions from autumn [79].

In contrast, neighboring user-based explanations calculate the similarity between users by comparing their ratings on shared items. This approach identifies "neighboring users"—a term broadly encompassing individuals with similar tastes, friends [80, 81], or other relevant social connections—and predicts preferences for unknown items based on these neighbors' preferences. User ratings and social information play a central role in generating such recommendations and their explanations, often leveraging highly rated items as exemplars to justify suggestions [82]. Commonly, UBCF techniques or K-Nearest Neighbors (KNN) algorithms are employed to facilitate these recommendations

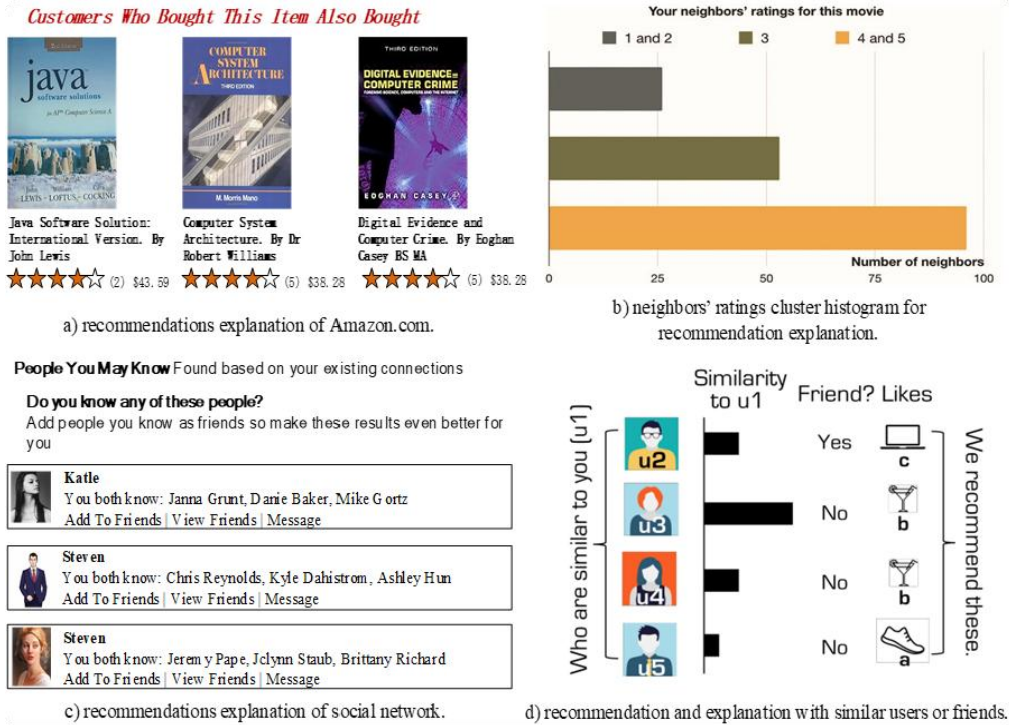

a) recommendations explanation of Amazon.com.

b) neighbors' ratings cluster histogram for recommendation explanation.

c) recommendations explanation of social network.

d) recommendation and explanation with similar users or friends.

**Fig. 6 User-based explanation examples**

In the context of user-based recommendation systems, explanations are frequently communicated through natural language descriptions, as exemplified in Fig. 6 (a & c). These early implementations, such as those seen on platforms like Amazon and Facebook, provide users with personalized suggestions based on implicit and explicit feedback. To improve the interpretability of these recommendations, several studies have explored the integration of textual explanations with visual elements. For instance, Fig. 6b presents a histogram of ratings from neighboring users, which allows users to visually assess the preferences of others who share similar tastes. This kind of visualization helps clarify the rationale behind recommendations by providing a clearer understanding of how a given suggestion is linked to others' interactions with the same items. Additionally, Fig. 6d illustrates how the preferences of a user's friends or social connections can be visualized to further enhance transparency in recommendation systems [9, 83]. By incorporating these visual aids, recommendation systems can offer more transparent and comprehensible explanations, ultimately improving user trust and engagement with the system.

Schaffer et al. [84] introduced an interactive movie recommender system employing collaborative filtering algorithms, as illustrated in Fig. 7. The interface sequentially displays the user's movie rating history, the Top-K Similar Users, and the final list of Top-N Recommendations from left to right.

The system is highly interactive, allowing users to add, delete, or modify their rated movies, with recommendations updating in real time. Moreover, it enhances recommendation interpretability through visual representation: when users interact with elements within the interface—such as movie ratings or recommendation results—dark blue lines dynamically appear, intuitively illustrating the provenance of the recommendations. These lines indicate the relationships between the recommended movies, similar users, and their rating data, thereby enabling users to comprehend the underlying rationale of the recommendations.

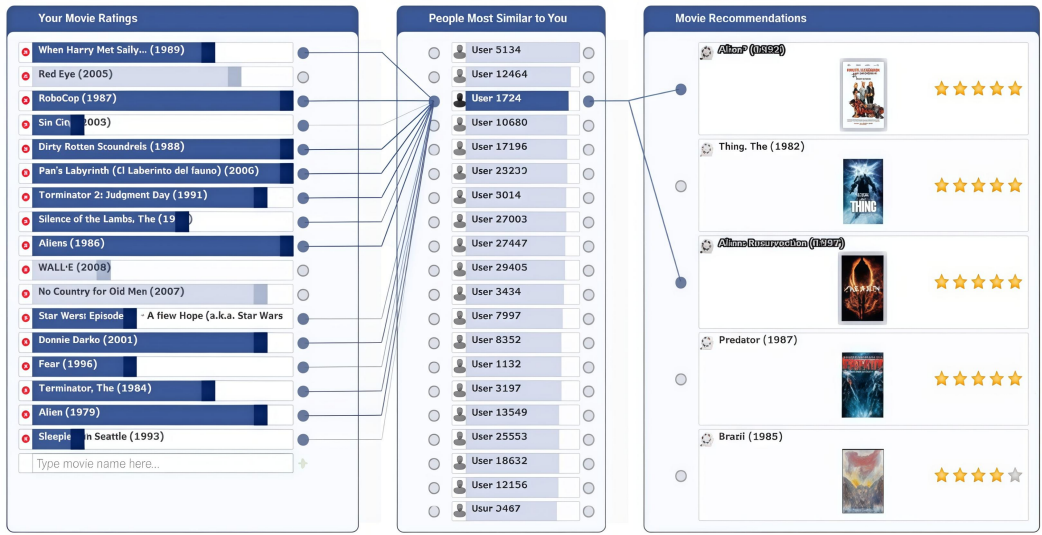

**Fig. 7 Interactive and explainable movie recommender system [84]**

Stetlin et al. [85] proposed LinkedVis (Fig. 8), an interactive visual recommender system that integrates social and semantic knowledge to offer career recommendations via the LinkedIn API. This system employs a collaborative approach to identify professionals with similar career trajectories and provides personalized suggestions for companies and roles.

The interface comprises three primary sections: (1) the left panel presents the user's profile information, including educational background, degrees, affiliated companies, and professional skills; (2) the central panel visualizes a network of connections, representing professionals with analogous career paths; and (3) the right panel offers career recommendations based on analyzed patterns, suggesting potential companies and roles. To enhance transparency and interpretability, the system employs a node-link diagram to depict relationships among user preferences, professional experiences, and recommendation outcomes. By enabling users to explore suggested career trajectories and their associated professional networks, LinkedVis facilitates data-driven decision-making in career development. The system's interactive nature allows users to refine recommendations based on personal preferences, fostering a more tailored and dynamic career exploration process.

Both systems depicted in Figs. 7 and 8 employ a node-link diagram to visualize the preferences or experiences of similar users, facilitating the generation of recommendations.

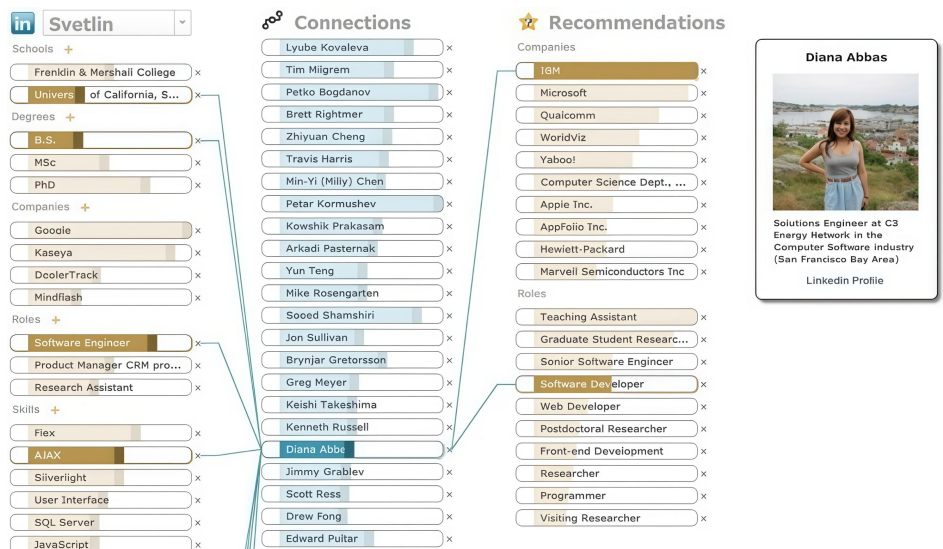

**Fig. 8 Interactive career recommender system – LinkedVis [85]**

## 4.2    Item-based Explanations

Item-based explanations typically inform users that the recommended items are likely to appeal to them, either due to frequent co-occurrence with previously interacted items or based on their similarity. These explanations often present a list of items the user has purchased, commented on, or browsed, emphasizing shared attributes such as name, function, utility, or brand. The three primary recommendation algorithms used for item-based explanations are content-based, association rule-based, and IBCF.

For association rule-based recommendations, the interpretability derives from the high co-occurrence frequency between items.

In contrast, the interpretability of content-based and IBCF methods is rooted in item similarity. For example, Fig. 9 (a & b) demonstrates Netflix's "Why recommend this movie?" feature, where the system highlights films that share similarities with those highly rated by the user in the past. This strategy helps users understand the rationale behind the recommendation, showcasing patterns in their viewing history [8]. Similarly, Amazon leverages the user's purchase history to anchor new suggestions. As shown in Fig. 9c, previously bought items are used as reference points to propose related products, thereby ensuring the recommendations align with the user's preferences [18]. Geo-system platforms adopt a similar logic, as illustrated in Fig. 9d, by recommending new locations based on previous visits. The explanation might state, "you have visited five similar places before," indicating that the recommendation is grounded in past user behavior. This approach not only clarifies the reasoning behind the suggestion but also strengthens user trust in the system by making the recommendations more personalized and relevant [86].

While item-based explanations are often presented in a straightforward manner, typically through natural language descriptions, they may not fully enable users to discern subtle differences and similarities between recommended items or facilitate detailed comparisons. This limitation has led researchers to explore the use of visual aids to improve the clarity and depth of these explanations.

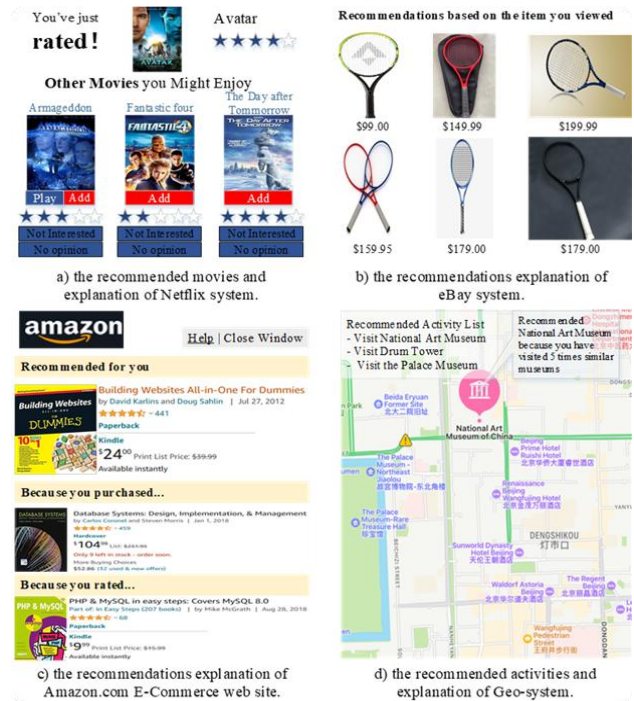

a) the recommended movies and explanation of Netflix system.

b) the recommendations explanation of eBay system.

c) the recommendations explanation of Amazon.com E-Commerce web site.

d) the recommended activities and explanation of Geo-system.

**Fig. 9 Item-based explanation examples**

A notable example is the introduction of "Introspective Views" by Bakalov et al. [87], as depicted in Fig. 10.

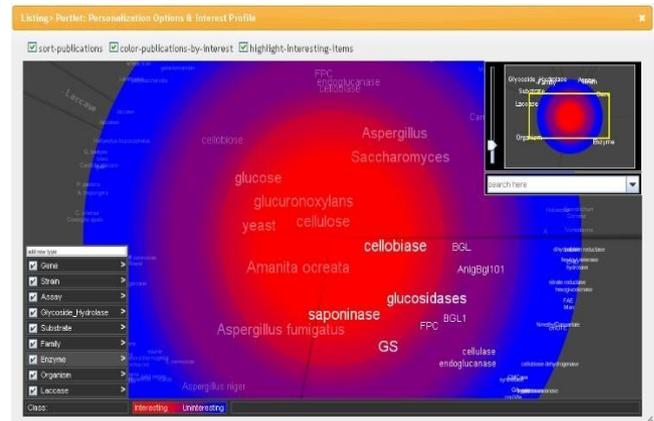

**Fig. 10 Introspective Views [87]**

This radial heatmap visualization provides a more intuitive and engaging way to explore item relationships by ranking items according to their engagement level and type. The visualization uses a central "hot zone" to display items that have garnered the most attention or interaction, whereas items with lower engagement are positioned in the peripheral "cold zone." This spatial arrangement allows users to easily identify high-engagement items and understand the relative significance of each item within the context of user interactions. By incorporating this radial visualization, Introspective Views enhance the interpretability of item-based recommendations, making it easier for users to explore and compare items based on their relevance and popularity. The dynamic nature of the visualization helps users gain deeper insights into the recommendation system's rationale and fosters a more user-centered experience.

### 4.3 Feature-based Explanations

Feature-based explanations emphasize how a recommended item aligns with a user's preferences by highlighting specific attributes. These explanations present key item features that user may find appealing, offering detailed and diverse insights. XRS often rely on sophisticated algorithms to analyze user preferences at the feature level. These range from basic content-based methods to advanced approaches such as explicit factor models, deep learning, knowledge graph-based techniques, and causal inference-based methods. Additionally, post-hoc explanation techniques are sometimes employed to provide detailed feature-level interpretations.

Feature-based explanations can be presented in various formats, including graphical labels, word clouds, template-based explanations, and sentence generation. This chapter provides a comprehensive overview of these explanation types alongside the supporting algorithms.

#### 4.3.1 Graphical labels explanation

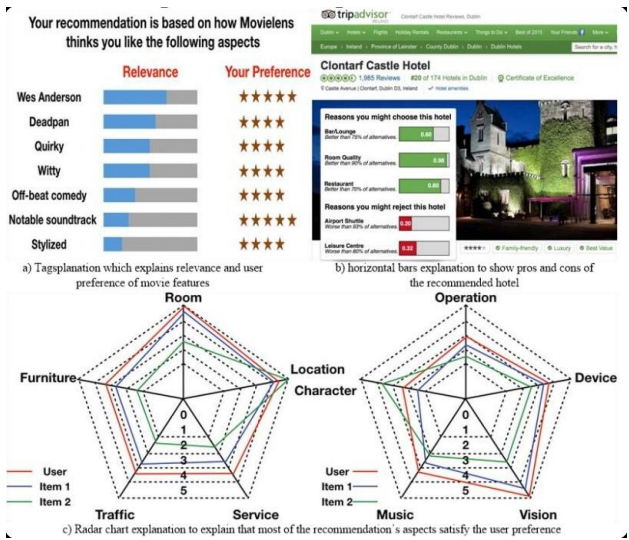

**Fig. 11 Feature-based explanation with Graphical labels**

This type of explanation aims to visually represent textual features through graphical dimensions and quantify them based on relevance or user preferences. It enables users to better understand and compare the features. For example, Tagsplanation, as shown in Fig. 11a [24], uses a content-based recommendation algorithm to evaluate the similarity between movies and users' preferences for specific tags. It adopts movie tags as features to generate recommendations and explanations, explaining to users why the recommendation is made by displaying their preferences for relevant features. Hou et al. propose the Aspect-based Matrix Factorization (AMF) model, which enhances rating prediction accuracy by incorporating aspect-based auxiliary information. They introduce two metrics: User Aspect Preference (UAP), which quantifies user preference for specific aspects, and Item Aspect Quality (IAQ), which measures review sentiment on an aspect. Using UAP and IAQ, a radar chart is employed to explain why a user selects a particular item, as shown in Fig. 11c [27]. Muhammad et al. proposed an approach for generating detailed and compelling explanations in recommender systems, based on opinions mined from user-generated reviews. By using a bar chart, the explanations highlight the features of a recommended item that matter most to the user and relate them to other recommendation alternatives, as shown in Fig 11b. These charts highlight the aspects that perfectly satisfy the target user.

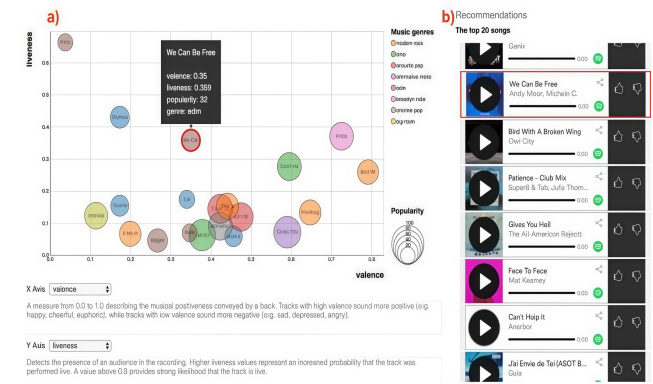

**Fig. 12 Combus [88]**

'ComBus' [88] employs a scatterplot to display recommended songs, using two specified audio features as the axes. The songs are presented as bubbles, with varying sizes and colors indicating their popularity scores and genres, respectively (Fig. 12).

Graphical labels explanations also include feature importance techniques in post-hoc methods. For example, Zafar et al. [89] proposed a deterministic version of LIME, replacing random perturbation with a structured approach. Agglomerative hierarchical clustering (HC) groups the training data, and KNN selects the relevant cluster for the instance being explained. As shown in Fig. 13, red bars represent negative coefficients and green bars represent positive coefficients in the linear regression model. Positive coefficients indicate a positive correlation between dependent and independent variables, while negative coefficients indicate a negative correlation. This method enhances interpretability by providing a clear and structured visualization of feature importance.

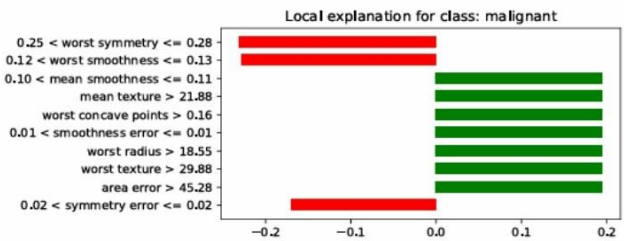

**Fig. 13 Explanations generated with DLIME [89]**

Similarly, the SHAP method provides valuable insights into the key factors influencing recommendation systems. Fig. 14 presents a comprehensive analysis of model behavior and effectiveness across the MovieLens and Amazon datasets. The upper panel illustrates feature importance, where the horizontal axis represents the normalized SHAP value scale (0-1) and the vertical axis lists the ranked features, A higher SHAP value indicates greater feature importance within the model. In Fig. 14, specific features, such as "Genre" and "User Activity," significantly affect the recommendations generated by the model.

While the lower panel depicts model performance variations across three key parameters: Number of Factors, Learning Rate, and Regularization. With the horizontal axis indicating the normalized hyperparameter values and the vertical axis showing the corresponding evaluation metric scores. Performance metrics exhibit pronounced sensitivity to the Number of Factors and Learning Rate, suggesting these parameters require meticulous tuning for optimal model configuration.

This comprehensive analysis provides actionable insights for optimizing recommendation systems through strategic feature selection and systematic hyperparameter tuning [90].

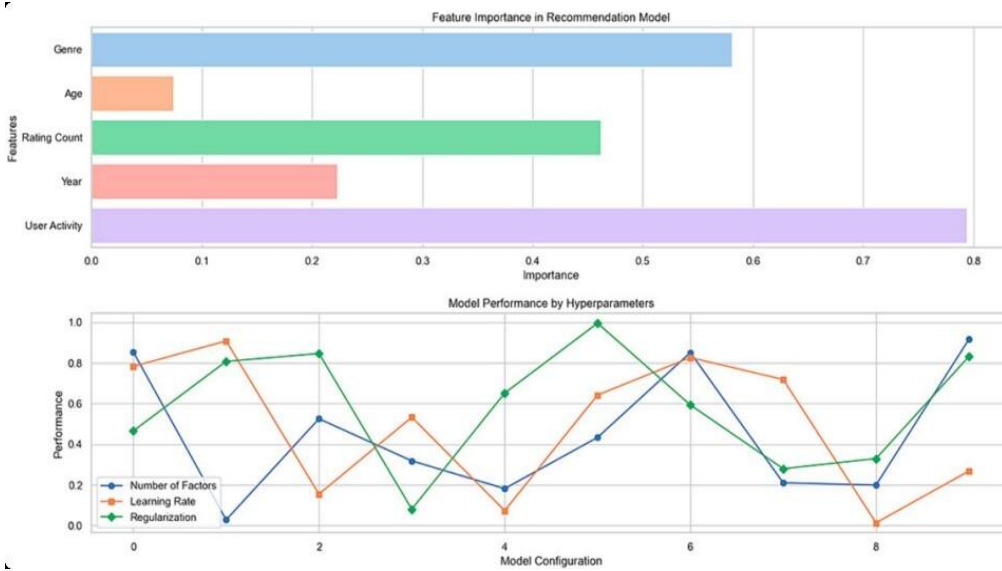

**Fig. 14 Explanations generated with SHAP [90]**

### 4.3.2 Word cloud explanation

These methods extract opinions on an item's tags or features from reviews and present them as word clouds for explanation, representing a content-analytics-based post-hoc explanation approach. Typically, the word size in the word cloud is proportional to the sentiment strength of the aspect, as shown in Fig. 15 [8]. Gedikli et al. further enhanced this approach by creating personalized, tag-based explanations that incorporated users' sentiments toward individual tags, which were visually expressed using different colors (Fig. 15a) [25].

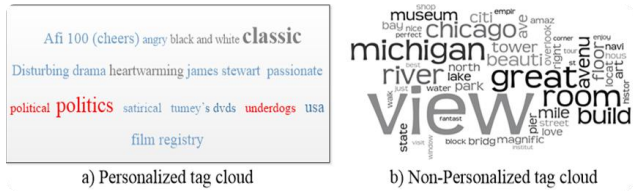

a) Personalized tag cloud          b) Non-Personalized tag cloud

**Fig. 15 Word cloud explanation** [8]

### 4.3.3 Template-based explanation

This approach uses predefined explanation sentence templates, which are then populated with different feature-related words to personalize explanations for individual users [28, 29], typically leverage post-hoc explanation methods to generate explanation.

For instance, Zhang et al. proposed the EFM to generate explainable recommendations by extracting explicit product features and user opinions through phrase-level sentiment analysis of user reviews. The model generates both recommendations and dis-recommendations based on these features, aligning them with the user's interests and learned latent factors. For example, it provides explanations like, "you might be interested in the feature on which this product performs well," with the feature selected via personalization algorithms (Fig. 16a & b) [10]. Similarly, dis-recommendations explain why an item may not be suitable, e.g., "you might be interested in the feature on which this product performs poorly." Building on this, Chen et al. proposed a mixture view of template-based explanations, which combines static specifications with sentiment-based features to construct more nuanced explanations, as shown in Fig. 16c [2].

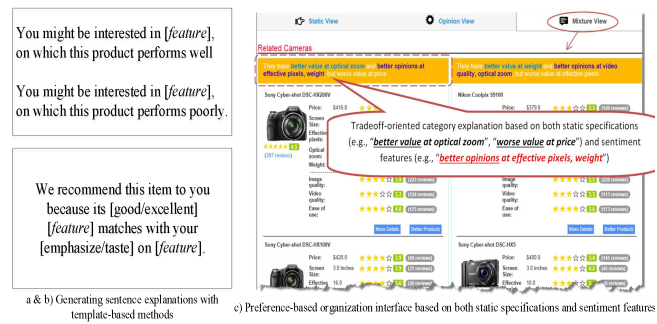

a & b) Generating sentence explanations with template-based methods

c) Preference-based organization interface based on both static specifications and sentiment features

**Fig. 16 Template-based explanation**

### 4.3.4 Generating sentence explanation

This method generates explanation sentences using natural language generation (NLG) techniques without relying on predefined templates. Typically, user reviews serve as the corpus, and related natural language processing techniques are employed to produce explanatory statements, making it a post-hoc explanation approach. For example, as shown in Fig. 17b [30], models trained on large-scale user reviews can generate reasonable review-like sentences as explanations. Additionally, Lin et al. enhanced explanations by labeling personalized features within an item's image for different users, as illustrated in Fig. 17a [28].

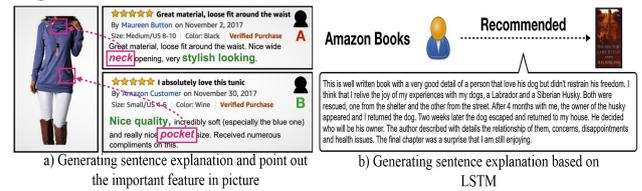

a) Generating sentence explanation and point out          b) Generating sentence explanation based on
the important feature in picture                                      LSTM

**Fig. 17 Generating sentence explanation**

Wang et al. introduced a reinforcement learning framework for explainable recommendations. Their model-agnostic framework can explain any recommendation system and flexibly adjust the quality of explanations based on specific application scenarios. Additionally, Shi et al. [73] leveraged the contextual understanding and language generation capabilities of large language models (LLMs) to create user-friendly explanations. Their approach combines reasoning over semantic relationships in knowledge graphs to form recommendations, followed by LLMs to generate post-hoc explanations.

## 4.4    Logical-based Explanation

Early on, logic-based explanations were primarily employed in expert decision support systems, such as those in medical domains. These explanations are grounded in knowledge- or rule-based algorithms built on domain ontologies or explicit rules, specifically designed to meet users' explicit needs. Recommendations in such systems are typically derived from predefined rules or ontologies, making the reasoning process relatively straightforward. However, explanations in these systems were often limited to system log recordings [91], which documented the sequence of rules used in decision-making. Since the advent of expert systems, automatically generated explanations have been regarded as critical for building user trust in system recommendations [92]. Nevertheless, these explanations are often difficult for non-experts to comprehend and are thus mainly utilized for system debugging or specialized decision-making tasks [93].

Recently, there has been a growing focus on providing users with explanations of the "recommendation process" to improve system transparency and trustworthiness. These reasoning processes are typically presented in graph formats. For example, Li et al. [58] propose a novel explainable framework targeting path-based recommendations, and design two counterfactual reasoning algorithms from both path representation and path topological structure perspectives, wherein the explainable weights of paths are learned to replace attention weights. As shown in Fig. 18, The first two paths are selected by Reinforcement Learning-based counterfactual reasoning, and the last one has a high explainable weight via attention training.

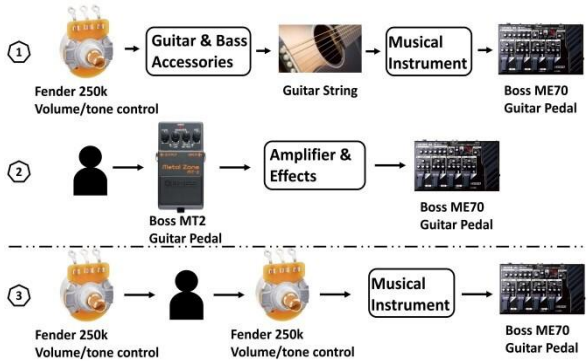

**Fig. 18 Path-based explanation [58]**

Due to the high level of background knowledge required to interpret such explanations, they are usually combined with other types of explanations to enhance accessibility and usability. As shown in Fig. 22 & 23 & 24.

## 4.5    Hybrid-content Explanation

In recent years, XRS have increasingly emphasized providing diverse and detailed explanations, leading to the development of hybrid explanation interfaces that combine multiple methods to cater to different audiences.

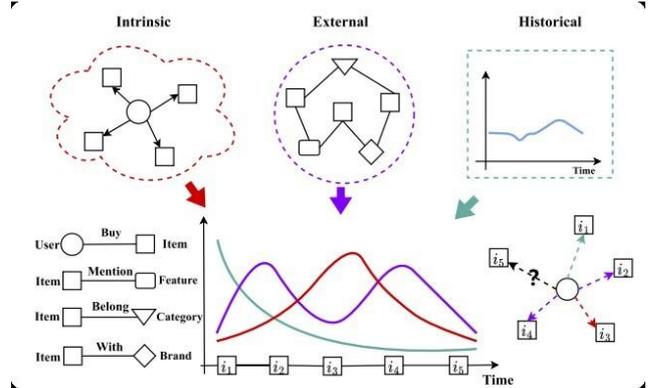

**Fig. 19 Explanations provided by FMHP [54]**

For example, Cui et al. [54] proposed the Factors Mixed Hawkes Process (FMHP), a model designed to enhance event-based incremental recommendations by incorporating three key influencing factors: intrinsic intensity (underlying personal factors driving user behavior), external intensity (external events influencing user actions), and historical intensity (the cumulative impact of past events). The model dynamically updates event formation as new events occur, effectively adapting recommendations to users' ongoing interactions. It also explains how these factors contribute to the generation of recommendations, as illustrated in Fig. 19, where the intensity of each factor changes over time, with varying contributions to triggering events. The explanations provided by FMHP span users, items, and features but are tailored primarily for expert users.

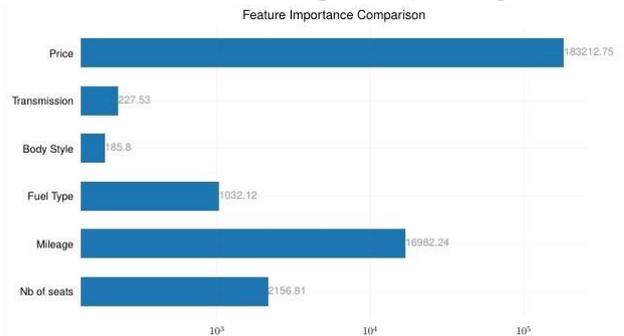

**Fig. 20 Feature importance explanation [94]**

Le et al. [94] introduced a multidimensional approach by integrating embedding-based and semantic-based models with ontology-based knowledge graphs. These graphs capture complex relationships between entities within a structured framework, enabling the generation of comprehensive and meaningful explanations. The explanations are designed for two distinct user groups: developers and end-users. For developers, a SHAP-based method (Fig. 20) quantifies feature importance during the training process, helping refine critical features to improve model performance.

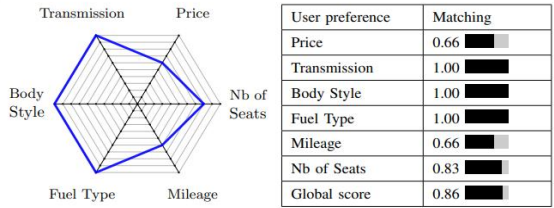

(a) Explanations using radar chart    (b) Explanations in tabular format

*The recommended item closely matches your preferences in several key aspects. It has a 66 % match with your proposed mileage. The vehicle body style, fuel type, and transmission align perfectly with your preferences, with a 100 % match for both categories. Additionally, the number of seats has an 83 % match, and the price aligns with your budget at a 66 % match. Overall, the item demonstrates an 86 % global match with your preferences, making it a highly suitable choice based on your requirements.*

(c) Explanations using natural language

**Fig. 21 A hybrid explanation of multiple feature levels [94]**

For end-users, feature-level explanations were provided in three formats as shown in Fig. 21: (a) a radar chart visualizes the alignment between features and user preferences, offering an intuitive representation of feature significance; (b) a tabular format presents a clear, comparative overview of features and their matching scores; and (c) natural language explanations provide a conversational and user-friendly narrative, eliminating the need for users to interpret complex visuals. These formats collectively enhance the transparency, usability, and interpretability of the recommendation process [94].

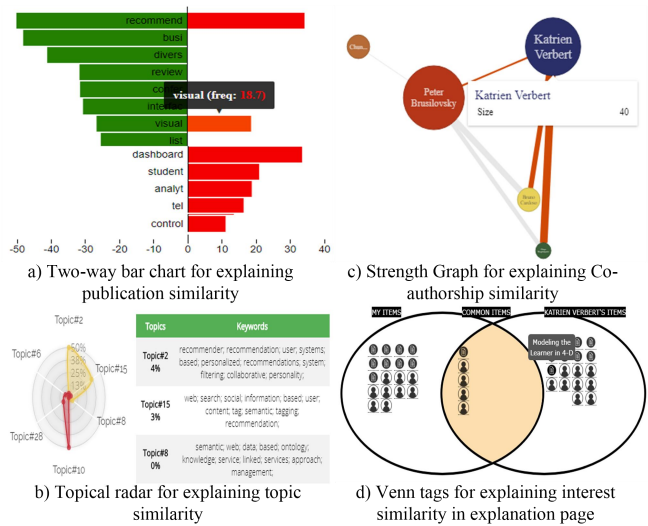

a) Two-way bar chart for explaining publication similarity

b) Topical radar for explaining topic similarity

c) Strength Graph for explaining Co-authorship similarity

d) Venn tags for explaining interest similarity in explanation page

**Fig. 22 Relevance Tuner+ [36]**

Tsai et al. [34, 36] introduced Relevance Tuner+, an explainable recommendation interface designed for the Conference Navigator platform, integrating four complementary visualization components, as illustrated in Fig. 22: a) Publication Similarity Analysis, which employs a two-way bar chart to represent cosine similarity between scholars' publications; b) Topic Similarity Visualization, where similarity is quantified by aligning research interests through a topic modeling approach (LDA); c) Co-authorship Network Analysis, which utilizes a strength graph to depict network proximity between authors; and d) Venn Tags for Interest Similarity, where similarity is assessed based on the number of co-bookmarked conference papers and shared co-authors within the Conference Navigator's social system. This interface allows users to fine-tune recommendation weights via interactive sliders while offering multi-dimensional explanations through these visualizations, effectively balancing system transparency with user control [36].

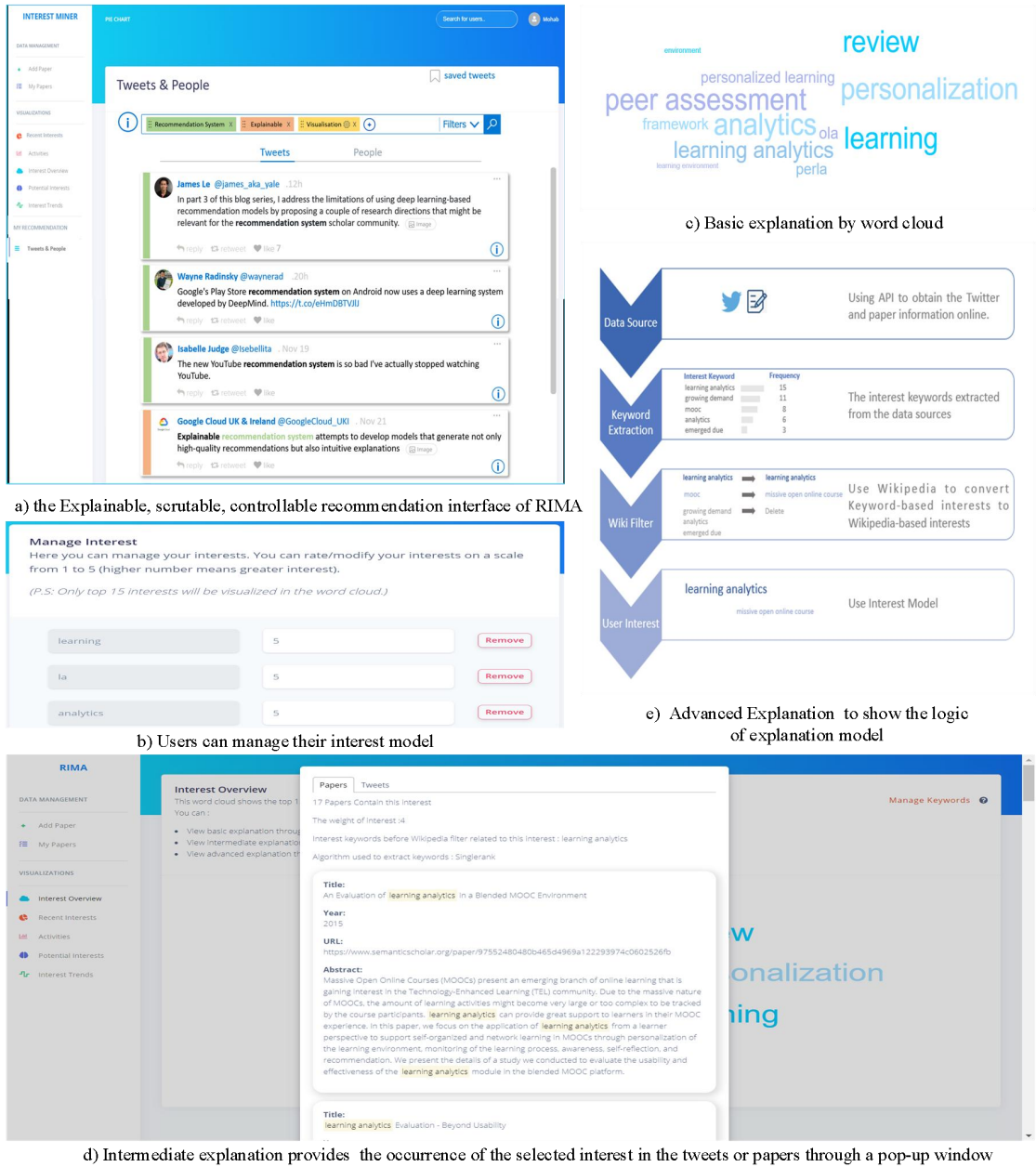

a) the Explainable, scrutable, controllable recommendation interface of RIMA

b) Users can manage their interest model

c) Basic explanation by word cloud

d) Intermediate explanation provides the occurrence of the selected interest in the tweets or papers through a pop-up window

e) Advanced Explanation to show the logic of explanation model

**Fig. 23 Recommendation and Interest Modeling Application (RIMA) [33]**

Guesmi et al. [33] introduced the Recommendation and Interest Modeling Application (RIMA), an explainable recommendation system that provides multi-level explanations through its interactive interface (Fig. 23). The system architecture comprises three core components: 1) a recommendation list page featuring explainable, scrutable, and controllable recommendations (Fig. 23a); 2) an interest management interface enabling users to modify and rate their preferences (Fig. 23b); and 3) a hierarchical explanation framework offering basic to advanced insights (Fig. 23c, d & e).

The explanation mechanism operates through three distinct levels: Basic level (Fig. 23c), Presents a tag cloud visualization of relevant publications when hovering over specific interests; Intermediate level (Fig. 23d), displays contextual occurrences of selected interests through pop-up windows, showing their appearance in both social media content (tweets) and academic literature (titles and abstracts); Advanced level (Fig. 23e), provides algorithmic transparency by demonstrating the computational logic behind interest inference, including feature weighting and decision pathways. This multi-level explanation framework effectively bridges the gap between system transparency and user comprehension, addressing diverse user needs while maintaining system usability.

a) top-10 recommended research and a line explanation

b) KG-path explanation for a particular recommendation

c) the organization explanation interface with items and features

**Fig. 24 Hybrid explanation example**

Chen et al. [2] proposed an organizational interface that presents the most popular item alongside 24 additional candidates. These candidates are grouped into four categories (k = 4) based on an organization selection and ranking algorithm. Each item is accompanied by a one-line explanation and detailed feature information (Fig. 24c), allowing users to compare and understand the relevance of the recommendations.

Purificato et al. [95] developed a Knowledge Graph-Based Recommender System for research paper suggestions, featuring two explainable interfaces. The first interface presents the top 10 recommended papers along with a concise explanation, highlighting the dominant subgraph similarity between the user and each suggested paper (Fig. 24a). To enhance transparency, users can further investigate the system's reasoning by tracing the path within the knowledge graph that links the user node to the research paper node, thereby facilitating a deeper comprehension of the recommendation process. As illustrated in Fig. 24b, the relationships among researchers, research interests, and research outputs serve as the basis for explaining the recommended papers.

These studies collectively demonstrate how hybrid explanations—combining multidimensional and multi-level interfaces—can effectively enhance system transparency, improve user understanding, and foster trust in recommender systems. By catering to both developers and end-users, such interfaces address the diverse needs of stakeholders while enabling greater control over the recommendation process.

# 5    Display Methods of XRS

The display method in XRS refers to the specific ways or means used to present recommendation explanations. Nunes and Jannach [15] categorized explanation presentations into natural language, visualizations, lists, logs, arguments, and other formats based on a static analysis of related literature. Mohamed et al. [16] suggested that the format of recommendation explanations can generally be classified into two categories: textual explanations and visual explanations. For presentation formats, we believe Mohamed's classification is more representative because lists, logs, or arguments are also displayed in text or visual forms. We categorize existing explanation displays into four types: textual explanation, visualization explanation, Hybrid-element explanation, and multimedia explanation. The latter two have emerged as explanation methods in recent years.

## 5.1    Textual Explanation

Textual explanations describe the reasons or logic behind recommendations using written language. Textual explanations were the earliest method adopted for presenting explanations. For instance, the explanation in Fig. 6a states "customers who bought this item also bought," and the explanation in Fig. 9b says "recommendations because you've viewed." Later, with the development of generative AI models, more complex generative text explanations began to emerge. For example, template-based generated sentence explanations, as shown in Fig. 16a, and NLG-based generative sentence explanations, as shown in Fig. 17b.

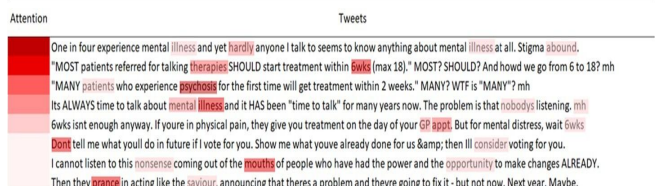

**Fig. 25 Explainability via visualization of attention score in MDHAN [96]**

In addition to template-based and sentence-generation methods, a new textual explanation approach has emerged in recent years. This method operates similarly to feature importance by highlighting segments of the original text that contribute most to the recommendation or prediction. For example, Zogan et al. [96] proposed an explainable model, Multi-Aspect Depression Detection with Hierarchical Attention Network (MDHAN), for automatic depression detection on social media. The model enhances explainability by incorporating a hierarchical attention mechanism at both the tweet-level and word-level, identifying the most influential tweets and words while capturing semantic sequence features from user timelines. By leveraging multi-aspect features and attention-based techniques, MDHAN provides interpretable predictions, allowing users to identify key patterns that contribute to its results, as illustrated in Fig. 25.

In the 2017 survey [15], the authors observed that text-based explanations were the most common presentation method, with the majority of approaches relying on textual formats to convey explanations to users. However, in more recent research, graphical and visualization techniques have increasingly become the dominant methods for presenting explanations.

## 5.2    Visualization Explanation

Visualization explanations offer users visualizations as a means of explanation. These visualizations can take the form of charts, images (either complete images or specific highlighted areas), or graphs. Visual explanations have the advantage of conveying more information than textual explanations while demanding less cognitive effort from the user to comprehend [97]. Mohamed et al. [16] conducted a systematic analysis of Visualization-based explanations, offering a comprehensive overview. They delineated that Visualization-based explanations encompass a variety of visualizations including node-link diagrams, bar charts, tag clouds, Venn diagrams, heat maps, tree maps, scatter plots, radar charts, pie charts, maps, and path graph. Notably, node-link diagrams, bar charts, and tag clouds emerge as the most prevalent choices. The utilization and application scenarios of these diverse visualization styles are consolidated in Table 1.

**Table 1. Utilization and application scenarios of visualization styles**

| Name | Primary Utilization and Application Scenarios | Examples |
|---|---|---|
| Node-link diagram | It is employed to illustrate the relationships between entities such as users, items, and features. This visualization diagram is particularly effective for network-based datasets, facilitating efficient path identification and topology exploration | Fig. 7 [84] Fig. 8 [85] |
| Bar chart | It is primarily used to compare the similarity between user preferences and recommendations. Additionally, it can compare differences in various attributes between different items and evaluate the importance of features. | Fig. 6b [9] Fig. 20 [94] Fig. 22a [36] |
| Tag/Word cloud | It is mostly used in Feature-based explanation, for summarizing or comparing features of different items | Fig. 15 [8] |
| Venn diagram | It is utilized to illustrate potential intersections between datasets, thus demonstrating the overlap between user preferences and recommendations. | Fig. 22d [36] |
| Heatmap | It is used to identify and summarize interesting or popular items. Additionally, it can be applied in attention models to highlight the importance of specific features or elements. | Fig. 10 [87] Fig. 25 [96] |
| Treemap | It can offer a hierarchical and space-efficient visual representation of data. | Ref [32] |
| Scatterplot | It is typically used to compare feature differences among recommended items along specific dimensions, such as the x-axis, y-axis, scatter size, or even color representing different features. | Fig. 12 [88] |
| Radar chart | It allows for the comparison of several attributes simultaneously, making it easier to understand the strengths and weaknesses of different items across multiple dimensions. | Fig. 11c [27] Fig. 21a [94] |
| Pie chart | It can offer a clear and intuitive visual representation of the distribution and proportions of various attributes associated with recommended items. | Ref [25] |
| World Map | World maps can visualize and contextualize location-based data, making recommendations more relevant and understandable to users. | Fig. 9d [86] |
| Path graph | It used to illustrate the pathways that generate recommendations, typically depicting the relationships between users, items, and features. | Fig. 18 [58] |

## 5.3    Hybrid-element Explanation

Hybrid elements-based explanations combine textual elements and multiple types of visualization styles to provide a comprehensive and intuitive understanding of the recommendations. By integrating text-based explanations with various visualizations, hybrid explanations can present different facets of the data cohesively and handle diverse data types more effectively (Fig. 22-24). Additionally, hybrid explanations can include interactive elements that allow users to explore data in greater depth. For example, hovering over a data point on a scatter plot could reveal detailed information about that recommendation, or users could input specific requirements to adjust the recommendations (Fig. 23) [33].

## 5.4    Multimedia Explanation

It is noteworthy that although there is no research explicitly confirming the effectiveness of utilizing multimedia formats such as video and audio to explain recommended items, many applications have already embraced the use of videos to showcase recommendation lists. This approach integrates visual and auditory elements to offer users a more intuitive presentation of recommended content, thereby enhancing their understanding and interest, ultimately achieving the objective of explaining recommendations to users [19, 20]. For instance, Netflix goes beyond static recommendation lists by incorporating video previews. While browsing recommended content, users can watch brief video previews of each program, aiding in their comprehension of the content. Similarly, YouTube presents recommended videos with thumbnail images on its recommendation and video playback pages, occasionally featuring dynamic thumbnails (short video clips or GIFs) to entice user clicks for viewing. Amazon Live enables brands and sellers to showcase products through live video streams, accompanied by a list of recommended products below the video. Users can watch live video streams to witness the actual use of products and directly purchase items from the recommended list. On the leading global e-commerce platform Taobao, there are over 100 million micro-videos, with hundreds of thousands of new micro-videos generated daily. Over 30 million users browse these micro-videos daily, resulting in an estimated total transaction value of millions of dollars. Therefore, e-commerce providers greatly benefit from developing effective micro-video recommendation systems to enhance user experience, engagement, and retention.

Some scholars have explored and developed recommendation systems utilizing video presentation formats. For instance, Guo et al. introduced a multi-modal video highlight detection task, integrating video-related textual information as supervisory signals. Additionally, they proposed a graph-based product perception model to address multi-modal video highlight detection in e-commerce contexts, aiming to capture the most appealing segments for display to consumers, thus enhancing product click-through rates (Fig. 26) [40].

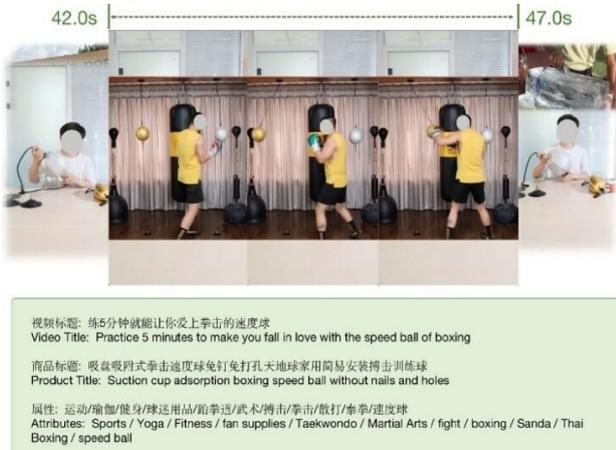

**Fig. 26 Video highlight explanation in E-commerce scene [40]**

Chen et al. developed a Sequential Multimodal Information Transmission Network (SEMI) that utilizes user behavior within the product domain to enhance micro-video recommendations. SEMI efficiently identifies pertinent items with multimodal characteristics across both micro-video and product domains, facilitating the depiction of user preferences and the distribution of micro-videos to potentially interested users (Fig. 27) [19].

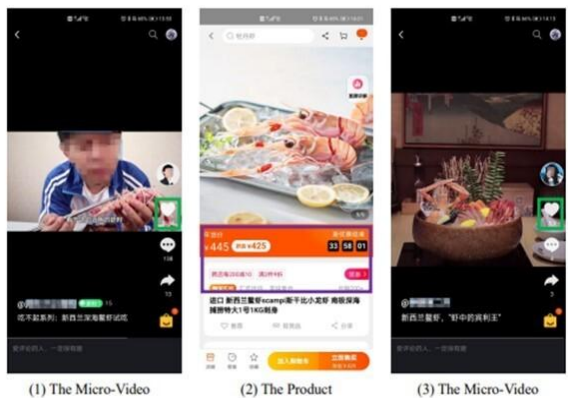

**Fig. 27 SEMI [19]**

Zhang et al. proposed a product-oriented video captioning framework, abbreviated as Poet, for generating video descriptions that narrate the preferred product features depicted in the video (Fig. 28) [20].

Therefore, although no scholars have explicitly proposed video-based explainable recommendation methods, videos hold significant potential due to their rich visual and auditory elements, which can intuitively convey product features and usage scenarios. Unlike text or static images, videos can dynamically illustrate complex relationships, demonstrate product functionality, and evoke emotional engagement, making them particularly effective in enhancing user understanding and trust. As the demand for more interactive and engaging explanations grows, video-based methods are poised to become a valuable addition to the field of explainable recommendation systems. Therefore, although no scholars have explicitly proposed video-based explainable recommendation methods, videos hold significant potential due to their rich visual and auditory elements, which can intuitively convey product features and usage scenarios. Unlike text or static images, videos can dynamically illustrate complex relationships, demonstrate product functionality, and evoke emotional engagement, making them particularly effective in enhancing user understanding and trust. As the demand for more interactive and engaging explanations grows, video-based methods are poised to become a valuable addition to the field of explainable recommendation systems.

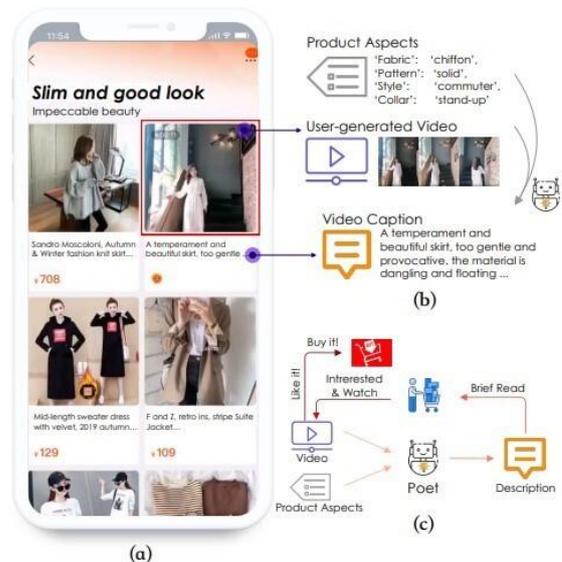

**Fig. 28 Poet [20]**

# 6 Evaluation Methods of XRS

The primary goal of XRS is to clarify the reasoning behind presented recommendations, assist users in forming stable preferences, and empower them to make informed, confident decisions [22]. Evaluating XRS primarily focuses on assessing whether explanations enhance users' acceptance of recommended results, thereby measuring the effectiveness of recommendation explanations. The evaluation methods can be broadly categorized into quantitative evaluation and qualitative evaluation [98], which some scholars further classify as objective evaluation and human-centered evaluation [99]. Quantitative evaluation primarily targets algorithms and can also extend to display content, while qualitative evaluation focuses on display methods and display content.

## 6.1 Quantitative evaluation

Quantitative evaluation methods can be divided into online and offline assessments. Online Evaluation requires access to a real-world system and involves human-computer interaction evaluation tasks. For instance, by comparing click-through rates, online evaluation measures the effectiveness of recommendation explanations. However, such evaluations can only be practically implemented in systems with a large user base, making this method both costly and challenging. Offline Evaluation is the most common approach in existing research. It measures the proportion of recommendations that can be explained by the explanation model, irrespective of explanation quality. Offline evaluations can be further divided into item-based and feature-based approaches.

Item-Based Offline Evaluation assesses the proportion of explainable items in the recommendation list. For example, Peake et al. [100] proposed the **Fidelity formula**, as Eq. (1):

$$\text{Fidelity} = \frac{|explainable\ items\ \cap\ recommended\ items|}{|recommended\ items|} \quad (1)$$

Additionally, Abdollahi et al. [101] introduced the metrics of **Mean Explainability Precision (MEP)** and **Mean Explainability Recall (MER)**.

**Explainability Precision (EP)**: The proportion of explainable items in a user's Top-N recommendation list. **Explainability Recall (ER)**: The proportion of explainable recommendations in the Top-N list relative to the total number of explainable items for the target user. MEP and MER are calculated as the average EP and ER values across all experimental users. Li et al. [58] further introduced the "Confidence" metric for explanation paths, which refers to the level of certainty in the generated paths. Higher confidence indicates more deterministic and stable explanation results, quantified by lower entropy in attention weights and path probabilities.

Feature-Based Offline Evaluation compares generated feature-level explanations with the actual features of recommended items, often applied in feature-mediated recommendation explanations. For example, Li et al. [102] proposed the following metrics. **Feature Matching Ratio (FMR)**: The proportion of generated explanation features that are actual features of the item; **Feature Coverage Ratio (FCR)**: The ratio of unique features included in all generated explanations to the total number of item features in the dataset; **Feature Diversity (FDIV)**: The degree of feature variation between any two generated explanations.

## 6.2 Qualitative evaluation

Qualitative evaluation methods primarily focus on understanding user experiences and perceptions of recommendation explanations. While these methods are essential for evaluating HCI-level aspects, they are inherently subjective and depend on user studies or surveys.

In practice, such evaluations are often guided by the Evaluation Criteria for Explanation proposed by Tintarev [22], which serve as the foundation for designing survey questionnaires and conducting user-centric investigations. Several studies [15, 16, 103] have summarized the key concepts and meanings of these criteria, as shown in Table 2. However, there has been limited exploration of specific measurement methods for evaluating the display of XRS interfaces. To address this gap, we incorporate scales and measures used in existing studies to provide a more comprehensive overview.

**Table 2 User evaluation criteria for recommendation explanation**

| Criteria | Application in studies |
|---|---|
| Transparency | This explanation helps me understand my predicted rating [10]. |
| | Do you think explanations help you better understand our system [11]? |
| | The system explained to me why the products were recommended [2]. |
| | I understood why the items were returned to me [2]. |
| | The interface helps me to understand why specific attendees were recommended [12]. |
| Scrutability | Allows me to give feedback on how well my preferences have been understood [101]. |
| Trust | The system can be trusted [2]. |
| | The trust is mainly assessed by three constructs: perceived competence, the intention to save effort, and the intention to return [23]. |
| | The interface helps me to improve my trust in the people recommendation result [36]. |
| Effectiveness | This explanation helps me determine how well I will like this movie [10]. |
| | Does the explanation help you know more about the recommended item [11]? |
| | This system helped me discover some useful info [2]. |
| | The system returned to me some good suggestions [2]. |
| | The shown scenario offers a good source of product information [101]. |
| | The interface helps me to explore various interesting people in the conference [36]. |
| Persuasiveness | The recommendation is convincing [102]. |
| Efficiency | This interface enabled me to compare different products very efficiently [23]. |
| | I easily found the information I was looking for [23]. |
| | Selecting a item using this system required too much effort [23]. |
| Satisfaction | This explanation helps me decide if this movie is right for my current mood [10]. |
| | Generally, are you satisfied with this recommendation [11]? |
| | Overall, I am satisfied with the system [2]. |
| | I like the people recommendation result from the system [36]. |
| Others | Do you think you get some idea about the recommended item? (Product knowledge) [11] |
| | How would you rate your knowledge about xxx? (Product knowledge) [11] |
| | I am very certain about what I need in respect of each attribute. (Preference certainty) [2]. |
| | The system helped me discover new products. (Perceived novelty) [2]. |
| | I feel in control of telling the recommender system what I want (Control) [35]. |
| | The layout of the interface is attractive and adequate (Adequacy) [35]. |
| | The shown scenario is enjoyable/pleasing/entertaining (Entertainment) [104]. |
| | The shown scenario is annoying/ frustrating/ irritating (Irritation) [104]. |
| | The consistency of the explainable model's results (Stability) [58]. |

As outlined in Table 2, most studies focus on evaluating users' perceptions of explanation interfaces along dimensions such as Transparency, Trust, Effectiveness, Efficiency, and Satisfaction. Additional dimensions, including Product Knowledge and Perceived Novelty, have also been explored in specific contexts [2, 11]. For example, some studies assess whether the system enables users to learn more about the recommended items or discover new products. Zimmermann et al. [104] expanded these evaluation criteria to include interface-specific attributes such as Entertainment, Adequacy, and Irritation. These metrics assess the overall design and emotional impact of the interface, offering insights into its usability and user experience. Notably, Li et al. were the first to propose the concept of explanation stability, which quantifies a qualitative concept by measuring the consistency of the explainable model's results [58].

Despite significant progress, certain criteria remain difficult to measure under experimental conditions, particularly Scrutability and Persuasiveness. To address these challenges, researchers have employed innovative experimental designs. Zhang et al. conducted A/B tests with real-world e-commerce users to evaluate the persuasiveness of tag-cloud explanations [12]. Their findings demonstrated the effectiveness of this explanation format in influencing user decisions. While Li et al. conducted a before-and-after experiment to measure differences in users' decision-making times and investigate the efficiency of explanations [2].

Currently, there is no standardized method for measuring Responsiveness in XRS. Ref. [15] distinguishes between explanations that are adapted to the user's current context and those that are not, emphasizing that more detailed explanations tend to be more responsive and informative. However, these detailed explanations can also impose higher cognitive demands on users, requiring a careful balance between informativeness and user effort. To address these gaps, future research could explore the following: 1) developing adaptive evaluation frameworks that assess explanations' ability to dynamically adjust to user preferences and contexts; 2) using advanced techniques like eye-tracking or interaction logs. Additionally, incorporating cognitive load metrics, such as task completion time or user-reported mental effort, could help evaluate the trade-off between informativeness and usability; 3) Longitudinal studies may also provide insights into how users' trust and satisfaction with explanations evolve over time. By addressing these challenges and expanding the scope of qualitative evaluation, researchers can develop more user-centric XRS that effectively balance transparency, usability, and personalization.

# 7     Conclusion and Future Direction

Given the growing interest in and rapid advancements of XRS, along with the limited research on HCI for XRS, this paper provides a comprehensive analysis, synthesis, and harmonization of existing taxonomies based on a detailed review of related studies and surveys. Furthermore, we examine HCI for XRS research from the perspectives of display content, display methods, and evaluation. The main contributions of this study are as follows:

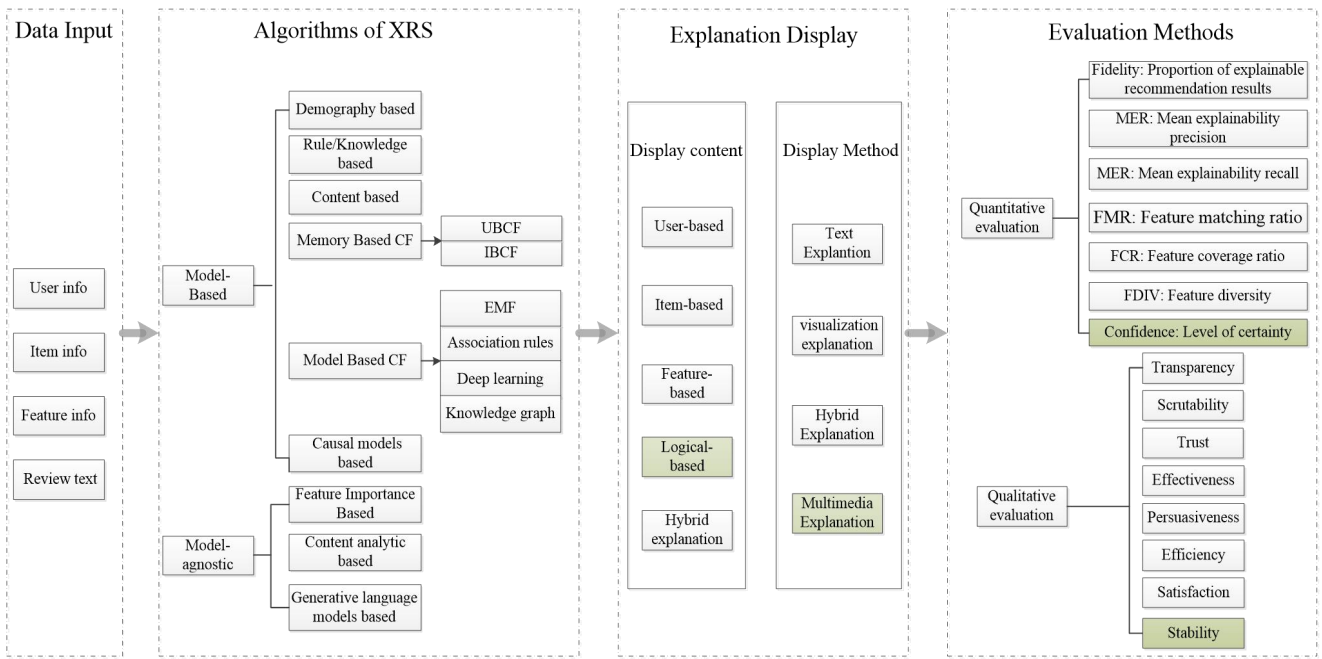

**Fig. 29 A comprehensive framework and categorization of XRS from a lifecycle perspective**

(1) We adopt a holistic perspective on the XRS lifecycle, systematically summarizing the technologies and methods used in XRS research and development. As illustrated in Fig. 29, the framework clearly indicates the type of recommendation algorithm, explanation model, display content and method employed by an XRS, as well as the evaluation dimensions used to assess the system. This approach addresses the challenge of categorizing explainable recommender systems due to the diversity and complexity of algorithmic models and implementation methods.

(2) For the first time, we propose that display methods for XRS should include Multimedia explanation techniques, especially video-based explanations which can enhance user understanding and engagement. In this regard, we also review video-related techniques, such as video extraction, summarization, and generation, within the context of recommender systems.

(3) We systematically summarize evaluation methods for XRS from both qualitative and

quantitative dimensions, offering a structured framework for assessing system effectiveness and usability.

Compared to the review [105], which proposed an ontological framework to define the requirements for XRS development, their model places greater emphasis on user needs and conceptual comprehensiveness. However, this focus results in models that are numerous and complex, while paying less attention to specific technical methods and content formats, especially given the rapid emergence of new technological directions in recent years. Based on our findings, researchers and practitioners can select appropriate data, algorithms, display formats, and evaluation methods to guide the systematic design, development, and testing of XRS. In the following section, we outline future research directions based on the conclusions drawn from this study.

(1) **Design and development of XRS across various domains** remains a timeless topic. In domain-specific recommender system research, balancing recommendation accuracy, diversity, and explainability continues to be a persistent challenge. For explanation display, a key consideration lies in determining how to present content and methods most effectively to highlight what users value the most. The effective integration of these elements can significantly enhance the quality of explanations. For instance, in academic recommendations, node-link graphs can visually depict relationships between scholars and research topics; in product recommendations, radar charts can compare product features; and in movie recommendations, video clips highlighting key moments or content can provide intuitive and engaging explanations.

(2) **Exploration and development of Multimedia-based methods for XRS.** The exploration of multimedia-based methods for XRS presents a promising opportunity to enhance explanation quality, user engagement, and system transparency. With their rich visual and auditory elements, videos can intuitively convey product details, illustrate complex relationships, and engage

users more effectively than text or static visuals. Several research paths can be pursued:

• **Video Creation and Generation for Explanations.** Development of algorithms to automatically generate concise, personalized video explanations tailored to user preferences and item features.

• **Integration of Video Summarization Techniques.** Application of video summarization methods to highlight specific aspects of products or services, such as key features, usage scenarios, or benefits.

While video-based methods offer significant potential, they also present notable technical and practical challenges:

• **Video Content Availability and Quality.** Ensuring access to high-quality video content relevant to recommended items is a critical challenge. Not all items may have suitable videos, creating gaps in explanation coverage.

• **Content Relevance and Summarization.** Selecting the most relevant video segments to align with user preferences and the recommendation context is complex. Automated video summarization and highlight detection require further refinement to produce concise and meaningful explanations.

• **Scalability.** Delivering personalized video explanations at scale demands substantial computational resources. Managing large volumes of video data and ensuring efficient streaming remain significant challenges.

(3) **For the qualified evaluation of XRS,** it is crucial to establish a comprehensive and systematic theoretical foundation along with a robust evaluation metric system. Existing frameworks, such as the seven main objectives of recommendation explanations [18], lack both theoretical depth and a well-structured research framework. Further efforts are needed to validate and refine evaluation metrics, standardize evaluation scales, and address key aspects such as responsiveness and cognitive load.

(4) **Cognitive load issues.** The three directions mentioned above all address the critical issue of cognitive burden. While recommender systems were originally designed to simplify decision-making and reduce users' cognitive load, explainable recommender systems risk reintroducing this burden through overly detailed or excessive explanations. This issue becomes even more pronounced with the increasing complexity of explanation content and formats, particularly in hybrid and multimedia explanation methods. We propose that future research should focus on the following three aspects to address this issue:

•**Prioritization of Relevant Information:** Hybrid and multimedia explanations should be designed to display only the most relevant and actionable information, tailored to the user's context and preferences.

•**Clear and Intuitive Presentation:** The display methods for explanations should employ clear visualizations, concise text, and well-structured multimedia formats to effectively simplify complex information and enhance user understanding.

•**Effective Measurement of Cognitive Burden:** Measuring cognitive burden is essential for evaluating system usability and efficiency. Future research should develop robust frameworks and metrics, incorporating indicators such as task completion time, user effort, error rates, and subjective assessments, to better understand and address user cognitive load in XRS.

These advancements will not only enrich HCI theories but also provide clearer guidance for the design, development, and improvement of XRS.

## References

[1] A. O. Nasraoui and S. Sanders, Mining Semantic Knowledge Graphs to Add Explainability to Black Box Recommender Systems, IEEE Access, vol. 7, pp. 110563-110579, 2019.

[2] L. Chen, D. Yan and F. Wang, User Evaluations on Sentiment-based Recommendation Explanations, ACM Trans. Interact. Intell. Syst., vol. 9, no. 4, pp. 1–38, 2019.

[3] C. Meske, E. Bunde, J. Schneider and M. Gersch, Explainable artificial intelligence: objectives, stakeholders, and future research opportunities, Information Systems Management, vol. 39, no. 1, pp. 53-63, 2022.

[4] K. Taha, P. D. Yoo, C. Yeun and A. Taha, Empirical and Experimental Perspectives on Big Data in Recommendation Systems: A Comprehensive Survey, Big Data Mining and Analytics, vol. 7, no. 3, pp. 964-1014, 2024.

[5] T. A. Bach, A. Khan, H. Hallock, G. Beltrao and S. Sousa, A systematic literature review of user trust in AI-enabled systems: An HCI perspective, International Journal of Human–Computer Interaction, vol. 40, no. 5, pp. 1251-1266, 2024

[6] E. Akhmetshin, G. Meshkova, M. Mikhailova, R. Shichiyakh, G. P. Joshi and W. Cho, Enhancing human computer interaction with coot optimization and deep learning for multi language identification, Scientific Reports, vol. 14, no. 1, pp. 22963, 2024.

[7] P. Shen, D. Wan and J. Li, How human–computer interaction perception affects consumer well-being in the context of online retail: From the perspective of autonomy, Nankai Business Review International, vol. 14, no. 1, pp. 102-127, 2023.

[8] Y. Zhang and X. Chen, Explainable Recommendation: A Survey and New Perspectives, FNT in Information Retrieval, vol. 14, no. 1, pp. 1–101, 2020.

[9] J. L. Herlocker, J. A. Konstan and J. Riedl, Explaining collaborative filtering recommendations, in Proceedings of the 2000 ACM conference on Computer supported cooperative work, Philadelphia Pennsylvania USA: ACM, pp. 241–250, 2000.

[10] Y. Zhang, G. Lai, M. Zhang, Y. Zhang, Y. Liu and S. Ma, Explicit factor models for explainable recommendation based on phrase-level sentiment analysis, in Proceedings of the 37th international ACM SIGIR conference on Research & development in information retrieval, Gold Coast Queensland Australia: ACM, pp. 83–92, 2014.

[11] N. Wang, H. Wang, Y. Jia and Y. Yin, Explainable Recommendation via Multi-Task Learning in Opinionated Text Data, in the 41st International ACM SIGIR Conference on Research & Development in Information Retrieval, Ann Arbor

MI USA: ACM, pp. 165–174. 2018.

[12] C. Y. Liu, W. Wu, S. Y. Wu, L. Yuan, R. Ding, F. h. Zhou and Q. H. Wu, Social-Enhanced Explainable Recommendation with Knowledge Graph, in IEEE Trans. Knowl. Data Eng., vol. 36, no. 2, pp. 840-853, 2024.

[13] D. Yu, Q. Li, X. M. Wang, Q. Li and G. d. Xu, Counterfactual explainable conversational recommendation, in IEEE Trans. Knowl. Data Eng, vol. 36, no. 6, pp. 2388-2400, 2024.

[14] J. S. Zhang, X. Chen, J. k. Tang, W. Q. Shao, Q. Y. Dai, Z. H. Dong and R. Zhang, Recommendation with Causality enhanced Natural Language Explanations, in Proceedings of the ACM Web Conference 2023, Austin TX USA: ACM, pp. 876–886, 2023.

[15] I. Nunes and D. Jannach. A systematic review and taxonomy of explanations in decision support and recommender systems, User Modeling and User-Adapted Interaction, vol. 27, pp. 393–444, 2017.

[16] M. A. Chatti, M. Guesmi and A. Muslim. Visualization for Recommendation Explainability: A Survey and New Perspectives, ACM Trans. Interact. Intell. Syst., vol. 14, no. 3, Article 19, 2024

[17] G. Friedrich and M. Zanker. A Taxonomy for Generating Explanations in Recommender Systems, AI Magazine, vol. 32, no. 3, pp. 90-98, 2011.

[18] A. Papadimitriou, P. Symeonidis and Y. Manolopoulos, A generalized taxonomy of explanations styles for traditional and social recommender systems, Data Min. Knowl. Disc., vol. 24, no. 3, pp. 555–583, 2012.

[19] C. Y. Lei, Y. Liu, L. Z. Zhang, G. X. Wang, H. H. Tang, H. Q. Li and C. Y. Miao, SEMI: A Sequential Multi-Modal Information Transfer Network for E-Commerce Micro-Video Recommendations, in Proceedings of the 27th ACM SIGKDD Conference on Knowledge Discovery & Data Mining, Virtual Event Singapore: ACM, pp. 3161–3171, 2021.

[20] S. Y. Zhang, Z. Q. Tan, J. Yu, Z. Zhao, K. Kuang, J. Liu, J. R. Zhou, H. X. Yang and F. Wu, Poet: Product-oriented Video Captioner for E-commerce, in Proceedings of the 28th ACM International Conference on Multimedia, Seattle WA USA: ACM, pp. 1292–1301, 2020.

[21] J. B. Schafer, J. Konstan and J. Riedl, Recommender systems in e-commerce, in Proceedings of the 1st ACM Conference on Electronic Commerce, Denver, CO, United States, pp. 158-166, 1999.

[22] N. Tintarev and J. Masthoff, A Survey of Explanations in Recommender Systems, in 2007 IEEE 23rd International Conference on Data Engineering Workshop, Istanbul, Turkey: IEEE, pp. 801–810, 2007.

[23] P. Pu and C. Li, Trust-inspiring explanation interfaces for recommender systems, Knowledge-Based Systems, vol. 20, no. 6, pp. 542-556, 2007.

[24] J. Vig, S. Sen, and J. Riedl, Tagsplanations: explaining recommendations using tags, in Proceedings of the 14th ACM international conference on Intelligent user interfaces, New York, NY, USA, pp. 47–56, 2009.

[25] F. Gedikli, D. Jannach, and M. Z. Ge, How should I explain? A comparison of different explanation types for recommender systems, International Journal of Human-Computer Studies, vol. 72, no. 4, pp. 367-382, 2014.

[26] K. I. Muhammad, A. Lawlor and B. Smyth, A Live-User Study of Opinionated Explanations for Recommender Systems, in Proceedings of the 21st International Conference on Intelligent User Interfaces, Sonoma California USA: ACM, pp. 256–260, 2016.

[27] Y. F. Hou, N. Yang, Y. Wu and P. S. Yu, Explainable recommendation with fusion of aspect information, World Wide Web, vol. 22, pp. 221–240, 2019.

[28] Y. J. Lin, P. J. Ren, Z. M. Chen, Z. C. Ren, J. Ma and M. D. Rijke, Explainable Outfit Recommendation with Joint Outfit Matching and Comment Generation, IEEE Trans. Knowl. Data Eng., vol. 32, no. 8, pp. 1502-1516, 2020.

[29] Y. Y. Tao, Y. L. Jia, N. Wang and H. N. Wang, The FacT: Taming Latent Factor Models for Explainability with Factorization Trees, in Proceedings of the 42nd International ACM SIGIR Conference on Research and Development in

Information Retrieval, Paris France: ACM, pp. 295–304, 2019.

[30] F. Costa, S. Ouyang, P. Dolog and A. Lawlor, Automatic Generation of Natural Language Explanations, in Companion Proceedings of the 23rd International Conference on Intelligent User Interfaces, Tokyo Japan: ACM, pp. 1–2, 2018.

[31] Z. H. Xie, S. Singh, J. McAuley and B. P. Majumder, Factual and informative review generation for explainable recommendation, in Proceedings of the AAAI Conference on Artificial Intelligence, vol. 37, no. 11, pp. 13816-13824, 2023.

[32] F. Du, C. Plaisant, N. Spring, K. Crowley and B. Shneiderman, EventAction: A Visual Analytics Approach to Explainable Recommendation for Event Sequences, ACM Trans. Interact. Intell. Syst., vol. 9, no. 4, pp. 21-52, 2019.

[33] M. Guesmi, M. A. Chatti, L. Vorgerd, S. Joarder, S. Zumor, Y. Q. Sun, F. Z. Ji and A. Muslim, On-demand Personalized Explanation for Transparent Recommendation, in Adjunct Proceedings of the 29th ACM Conference on User Modeling, Adaptation and Personalization, Utrecht Netherlands: ACM, pp. 246–252, 2021.

[34] C. H. Tsai and P. Brusilovsky, Explaining recommendations in an interactive hybrid social recommender, in Proceedings of the 24th International Conference on Intelligent User Interfaces, Marina del Ray California: ACM, pp. 391–396, 2019.

[35] Guesmi, M., Chatti, M. A., Joarder, S., Ain, Q. U., Alatrash, R., Siepmann, C., and Vahidi, T. Interactive Explanation with Varying Level of Details in an Explainable Scientific Literature Recommender System. International Journal of Human–Computer Interaction, vol. 40, no. 22, pp. 7248–7269, 2024

[36] C. H. Tsai and P. Brusilovsky, User Feedback in Controllable and Explainable Social Recommender Systems: A Linguistic Analysis, in CEUR Workshop Proceedings (IntRS '20), pp. 1-13, 2020.

[37] S. Naveed and J. Ziegler, Featuristic: An interactive hybrid system for generating explainable recommendations – beyond system accuracy, in CEUR Workshop Proceedings, pp. 14-25, 2020.

[38] Q. Q. Liang, X. L. Zheng, Y. Wang and M. Y. Zhu, O3ERS: An explainable recommendation system with online learning, online recommendation, and online explanation, Information Sciences, vol. 526, pp. 94-115, 2021.

[39] D. Jannach, S. Naveed and M. Jugovac, User Control in Recommender Systems: Overview and Interaction Challenges, in: Bridge, D., Stuckenschmidt, H. (eds) E-Commerce and Web Technologies. EC-Web 2016. Lecture Notes in Business Information Processing, vol. 278, pp. 21-28, 2017.

[40] Z. Y. Guo, Z. Zhao, W. K. Jin, D. Z. Wang, R. T. Liu and J. Yu, TaoHighlight: Commodity-Aware Multi-Modal Video Highlight Detection in E-Commerce, IEEE Trans. Multimedia, vol. 24, pp. 2606-2616, 2022.

[41] A. Sarkar, D. Vijaykeerthy, A. Sarkar and V. Balasubramanian, A Framework for Learning Ante-hoc Explainable Models via Concepts, 2022 IEEE/CVF Conference on Computer Vision and Pattern Recognition, 10276-10285, 2022.

[42] M. J. Pazzani, A framework for collaborative, content-based and demographic filtering, Artificial Intelligence Review, vol. 13, no. 5, pp. 393-408, 1999.

[43] X. W. Zhao, Y. Guo, Y. He, H. Jiang, Y. Wu and X. Li, We know what you want to buy: a demographic-based system for product recommendation on microblogs, KDD '14: Proceedings of the 20th ACM SIGKDD International Conference on Knowledge Discovery and Data Mining, pp. 1935-1944, 2014.

[44] P. Do, K. Nguyen, T. N. Vu, T. N. Dung and T. D. Le, Integrating knowledge-based reasoning algorithms and collaborative filtering into e-learning material recommendation system, In Future Data and Security Engineering: 4th International Conference, FDSE 2017, Ho Chi Minh City, Vietnam, pp. 419-432, 2017.

[45] Q. Guo, F. Zhuang, C. Qin, H. Zhu, X. Xie, H. Xiong and Q. He, A survey on knowledge graph-based recommender systems. IEEE

Transactions on Knowledge and Data Engineering, vol. 34, no. 8, pp. 3549-3568, 2020.

[46] G. Sunandana, M. Reshma, Y. Pratyusha, M. Kommineni and S. Gogulamudi, Movie recommendation system using enhanced content-based filtering algorithm based on user demographic data, International Conference on Communication and Electronics Systems (ICCES), Coimbatre, India, pp. 1-5, 2021.

[47] T. K. Nadar, S. S. Athulya, S. S. Rahul and K. Meenakshi, Job Recommendation by Content Filtering using TF-IDF and Cosine Similarity, 2024 8th International Conference on I-SMAC (IoT in Social, Mobile, Analytics and Cloud) (I-SMAC), Kirtipur, Nepal, pp. 1585-1590, 2024.

[48] M. J. Pazzani and D. Billsus, Content-based recommendation systems, In: The Adaptive Web. Springer, pp. 325–341, 2007.

[49] M. Hasan, S. Ahmed, M. A. I. Malik and S. Ahmed, A comprehensive approach towards user-based collaborative filtering recommender system, 2016 International Workshop on Computational Intelligence, Dhaka, Bangladesh, pp. 159-164, 2016.

[50] B. Sarwar, G. Karypis, J. Konstan and J. Riedl, Item-based collaborative filtering recommendation algorithms. In Proceedings of the 10th international conference on World Wide Web (WWW '01). Association for Computing Machinery, New York, NY, USA, pp. 285–295, 2001.

[51] P. Kumar, K. N. Manisha and M. Nivetha, Market Basket Analysis for Retail Sales Optimization, 2024 Second International Conference on Emerging Trends in Information Technology and Engineering, Vellore, India, pp. 1-7, 2024.

[52] W. Lin, S. A. Alvarez and C. Ruiz, Efficient adaptive-support association rule mining for recommender systems, Data Mining & Knowledge Discovery, vol. 6, no. 1, pp. 83-105, 2002.

[53] Y. A. Ünvan, Market basket analysis with association rules, Communications in Statistics-Theory and Methods, vol. 50, no. 7, pp. 1615-1628, 2021.

[54] Z. Cui, X. Sun, L. Pan, S. Liu and G. Xu, Event-based incremental recommendation via factors mixed Hawkes process. Information Sciences, vol. 639, pp. 119007, 2023.

[55] C. Zou and Z. Chen, Joint latent factors and attributes to discover interpretable preferences in recommendation, Information Sciences, vol. 505, pp. 498-512, 2019.

[56] S. Zhang, H. Chen, H. Yang, X. Sun, P. S. Yu and G. Xu, Graph masked autoencoders with transformers, arXiv preprint arXiv:2202.08391, 2022.

[57] A. L. Zanon, L. C. D. da Rocha and M. G. Manzato, Model-agnostic knowledge graph embedding explanations for recommender systems, World Conference on Explainable Artificial Intelligence, Cham: Springer Nature Switzerland, pp. 3-27, 2024.

[58] Y. Li, X. Sun, H. Chen, S. Zhang, Y. Yang and G. Xu, Attention Is Not the Only Choice: Counterfactual Reasoning for Path-Based Explainable Recommendation, in IEEE Transactions on Knowledge and Data Engineering, vol. 36, no. 9, pp. 4458-4471, 2024.

[59] F. Aghaeipoor, M. Sabokrou and A. Fernández, Fuzzy Rule-Based Explainer Systems for Deep Neural Networks: From Local Explainability to Global Understanding, in IEEE Transactions on Fuzzy Systems, vol. 31, no. 9, pp. 3069-3080, 2023.

[60] Z. Niu, G. Zhong and H. Yu, A review on the attention mechanism of deep learning, Neurocomputing, vol. 452, pp. 48-62, 2021.

[61] X. Wu, H. Wan, Q. Tan, W. Yao, and N. Liu, DIRECT: Dual Interpretable Recommendation with Multi-aspect Word Attribution, ACM Transactions on Intelligent Systems and Technology, vol. 15, no. 5, pp. 1-21, 2024.

[62] M. Klimo, J. Kopčan and L. Králik, Explainability as a Method for Learning From Computers, in IEEE Access, vol. 11, pp. 35853-35865, 2023.

[63] S. Zhang, Z. Jiang, J. Yao, F. Feng, K. Kuang and Z. Zhou, Causal Distillation for Alleviating Performance Heterogeneity in Recommender Systems, in IEEE Transactions on Knowledge and Data Engineering, vol. 36, no. 2, pp. 459-474, 2024.

[64] Y. Ding, Y. Sun and J. Feng, The Application of Causal Inference Algorithms in Federated Recommender Systems, in IEEE Access, vol. 12,

pp. 29748-29758, 2024.

[65] M. Jian, Y. Bai, X. Fu, J. Guo, G. Shi and L. Wu, Counterfactual Graph Convolutional Learning for Personalized Recommendation, ACM Transactions on Intelligent Systems and Technology, vol. 15, no. 4, pp. 67-87, 2024.

[66] R. Cai, F. Wu, Z. Li, J. Qiao, W. Chen, Y. Hao, and H. Gu, REST: Debiased Social Recommendation via Reconstructing Exposure Strategies, ACM Transactions on Knowledge Discovery from Data, vol. 18, no. 2, pp. 38-62, 2024.

[67] D. Slack, A. Hilgard, S. Singh and H. Lakkaraju, Reliable post hoc explanations: Modeling uncertainty in explainability, Advances in neural information processing systems, vol. 34, pp. 9391-9404, 2021.

[68] A. Madsen, S. Reddy and S. Chandar, Post-hoc interpretability for neural nlp: A survey. ACM Computing Surveys, vol. 55, no. 8, pp. 1-42, 2022.

[69] Hooshyar and Y. Yang, Problems With SHAP and LIME in Interpretable AI for Education: A Comparative Study of Post-Hoc Explanations and Neural-Symbolic Rule Extraction, in IEEE Access, vol. 12, pp. 137472-137490, 2024.

[70] S. Ouyang and A. Lawlor, Improving Explainable Recommendations by Deep Review-Based Explanations, in IEEE Access, vol. 9, pp. 67444-67455, 2021.

[71] F. Zafari, I. Moser and T. Sellis, ReEx: An integrated architecture for preference model representation and explanation, Expert Systems with Applications, vol. 161, pp. 113706, 2020.

[72] J. Xie, F. Zhu, X. Li, S. Huang and S. Liu, Attentive preference personalized recommendation with sentence-level explanations, Neurocomputing, vol. 426, pp. 235-247, 2021.

[73] G. Shi, X. Deng, L. Luo, L. Xia, L. Bao, B. Ye, F. Du, S. Pan and Y. Li, Llm-powered explanations: Unraveling recommendations through subgraph reasoning, arXiv preprint arXiv:2406.15859, 2024.

[74] K. Balog, F. Radlinski and S. Arakelyan, Transparent, Scrutable and Explainable User Models for Personalized Recommendation, in Proceedings of the 42nd International ACM SIGIR Conference

on Research and Development in Information Retrieval, Paris France: ACM, pp. 265–274, 2019.

[75] C. Roberto, L. Coba, B. Wagner and T. R. Besold, A historical perspective of explainable Artificial Intelligence, Wiley Interdisciplinary Reviews: Data Mining and Knowledge Discovery, vol. 11, no. 1, pp. e1391, 2021.

[76] W. Li, W. Wang, W. Huang, M. Tian and R. Zhang, A Review of Explainable Recommendation, Journal of the China Society for Scientific and Technical Information, vol. 42, no. 07, pp. 870-882, 2023.

[77] Y. Wei, X. Wang, L. Nie, S. Li, D. Wang and T. S. Chua, Causal Inference for Knowledge Graph Based Recommendation, in IEEE Transactions on Knowledge and Data Engineering, vol. 35, no. 11, pp. 11153-11164, 2023.

[78] W. X. Zhao, S. Li, Y. He, L. Wang, J. R. Wen and X. Li, Exploring demographic information in social media for product recommendation. Knowledge and Information Systems, vol. 49, pp. 61-89, 2016.

[79] K. K and S. T, Enhanced Product Recommendations based on Seasonality and Demography in Ecommerce, 2020 2nd International Conference on Advances in Computing, Communication Control and Networking, Greater Noida, India, pp. 721-723, 2020.

[80] A. Sharma and D. Cosley, Do social explanations work? studying and modeling the effects of social explanations in recommender systems, in Proceedings of the 22nd international conference on World Wide Web, Rio de Janeiro Brazil: ACM, pp. 1133–1144, 2013.

[81] Y. Li, J. Liu and J. Ren, Social recommendation model based on user interaction in complex social networks, PloS one, vol. 14, no. 7, pp. e0218957, 2019.

[82] K. Ji and H. Shen, Jointly modeling content, social network and ratings for explainable and cold-start recommendation, Neurocomputing, vol. 218, pp. 1-12, 2016.

[83] H. Park, H. Jeon, J. Kim, B. Ahn and U. Kang, UniWalk: Explainable and Accurate Recommendation for Rating and Network Data, arXiv e-prints, 2017, Article arXiv:1710.07134.

[84] S. James, T. Hollerer and J. O'Donovan, Hypothetical recommendation: A study of interactive profile manipulation behavior for recommender systems, The twenty-eighth international flairs conference, 2015.

[85] S. Bostandjiev, J. O'Donovan and T. Höllerer, LinkedVis: exploring social and semantic career recommendations, in Proceedings of the 2013 international conference on Intelligent user interfaces, Santa Monica California USA: ACM, pp. 107–116, 2013.

[86] V. W. Zheng, Y. Zheng, X. Xie and Q. Yang, Collaborative location and activity recommendations with GPS history data, in Proceedings of the 19th international conference on World wide web, Raleigh North Carolina USA: ACM, pp. 1029–1038, 2010.

[87] F. Bakalov, M. J. Meurs, B. K. Ries, B. Sateli, R. Witte, G. Butler and A. Tsang, An approach to controlling user models and personalization effects in recommender systems, in Proceedings of the 2013 international conference on Intelligent user interfaces, Santa Monica California USA: ACM, pp. 49–56, 2013.

[88] Y. C. Jin, N. Tintarev and K. Verbert, Effects of Individual Traits on Diversity-Aware Music Recommender User Interfaces, in Proceedings of the 26th Conference on User Modeling, Adaptation and Personalization, Singapore Singapore: ACM, pp. 291–299, 2018.

[89] M. R. Zafar and N. M. Khan, DLIME: A deterministic local interpretable model-agnostic explanations approach for computer-aided diagnosis systems, arXiv preprint arXiv:1906.10263, 2019.

[90] J. Govea, R. Gutierrez and W. Villegas-Ch, Transparency and precision in the age of AI: evaluation of explainability-enhanced recommendation systems. Frontiers in Artificial Intelligence, vol. 7, pp. 1410790, 2024.

[91] A. Basu and R. Ahad, Using a relational database to support explanation in a knowledge-based system, IEEE Trans. Knowl. Data Eng., vol. 4, no. 6, pp. 572-581, 1992.

[92] L. R. Ye and P. E. Johnson, The impact of explanation facilities on user acceptance of expert systems advice, Mis Quarterly, pp. 157-172. 1995.

[93] H.A Güvenir and N. Emeksiz, An expert system for the differential diagnosis of erythemato-squamous diseases, Expert Systems with Applications, vol. 18, no. 1, pp. 43-49, 2000.

[94] N. L. LE, M. H. ABEL and GOUSPILLOU P, Combining Embedding-Based and Semantic-Based Models for Post-Hoc Explanations in Recommender Systems, 2023 IEEE International Conference on Systems, Man, and Cybernetics, Honolulu, Oahu, HI, USA, pp. 4619-4624, 2023.

[95] E. Purificato, B. A. Manikandan, P. V. Karanam, M. V. Pattadkal and E. W. D. Luca, Evaluating Explainable Interfaces for a Knowledge Graph-Based Recommender System, in Proceedings of the CEUR Workshop Proceedings (RecSys 2021), pp. 73-88, 2021.

[96] H. Zogan, I. Razzak, X. Wang, S. Jameel and G. Xu, Explainable depression detection with multi-aspect features using a hybrid deep learning model on social media. World Wide Web, vol. 25, no. 1, pp. 281-304, 2022.

[97] W. Colin, Information visualization: perception for design, Morgan Kaufmann, pp. 10-11, 2019.

[98] K. Balog and F. Radlinski, Measuring Recommendation Explanation Quality: The Conflicting Goals of Explanations, in Proceedings of the 43rd International ACM SIGIR Conference on Research and Development in Information Retrieval, Virtual Event China: ACM, pp. 329–338, 2020.

[99] P. Kouki, J. Schaffer, J. Pujara, J. O'Donovan and L. Getoor, Personalized explanations for hybrid recommender systems, in Proceedings of the 24th International Conference on Intelligent User Interfaces, Marina del Ray California: ACM, pp. 379–390, 2019.

[100] G. Peake and J. Wang, Explanation mining: post hoc interpretability of latent factor models for recommendation systems, KDD '18: Proceedings of the 24th ACM SIGKDD International Conference on Knowledge Discovery & Data Mining, ACM, pp. 2060–2069, 2018.

[101] B. Abdollahi and O. Nasraoui, Using explainability for constrained matrix factorization, In: Proceedings of the 11th ACM Conference on Recommender Systems. ACM, pp. 79–83, 2017.

[102] L. Li, Y. Zhang and L. Chen, Extra: Explanation ranking datasets for explainable recommendation, Proceedings of the 44th International ACM SIGIR conference on Research and Development in Information Retrieval, pp. 2463-2469, 2021.

[103] A. Vultureanu-Albişi and C. Bădică, A survey on effects of adding explanations to recommender systems, Concurrency and Computation: Practice and Experience, vol. 34, no. 20, pp. e6834, 2022.

[104] R. Zimmermann, D. Mora, D. Cirqueira, M. Helfert, M. Bezbradica, D. Werth, W. J. Weitzl, R. Riedl and A. Auinger, Enhancing brick-and-mortar store shopping experience with an augmented reality shopping assistant application using personalized recommendations and explainable artificial intelligence, Journal of Research in Interactive Marketing, vol. 17, no. 2, pp. 273–298, 2023.

[105] M. Caro-Martínez, G. Jiménez-Díaz and J. A. Recio-García, Conceptual modeling of explainable recommender systems: an ontological formalization to guide their design and development, Journal of Artificial Intelligence Research, vol. 71, pp. 557-589, 2021.

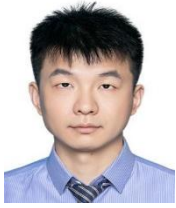

**Weiqing Li** received the Ph.D. degree in management science and engineering from Centre China Normal University, Wuhan, Hubei, China, in 2021. He is a teacher of the School of Economics and Management, Hubei University of Technology. His current research interests primarily focus on explainability in machine learning and explainable personalized recommendations. He has published eight journal articles on related topics.

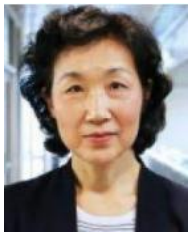

**Yue Xu (IEEE Member)** received the Ph.D. degree in computing from the University of New England, Armidale, NSW, Australia, in 2000. She is an Active Researcher in web intelligence and data mining. She is currently an Associate Professor with the School of Computer Science, Queensland University of Technology, Brisbane, QLD, Australia. She has published more than 180 refereed articles. Her current research interests include recommender systems, text mining, pattern and association mining, and user interest and behavior modeling.

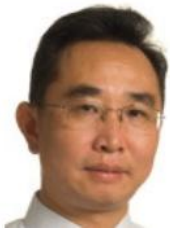

**Yuefeng Li (IEEE Member)** received the Ph.D. degree in computer science from Deakin University, Melbourne, VIC, Australia, in 2001. He is currently a Professor and the HDR Director of the School of Computer Science, Queensland University of Technology, Brisbane, QLD, Australia. He has published more than 190 refereed journals articles and conference papers. His work has received 5190 career citations (2418, since 2015) with an overall H-index of 34 and an i10 index of 101. He has published ten articles with more than 100 citations and three articles with more than 200 citations (Google Scholar 10/2021). He has demonstrable experience in leading large-scale research projects and achieved many established research outcomes that has been published and highly-cited in top data mining journals and conferences, such as the IEEE TRANSACTIONS ON KNOWLEDGE AND DATA ENGINEERING, IEEE ACCESS, ACM SIGKDD Conference on Knowledge Discovery and Data Mining, ACM International Conference on Information and Knowledge Management (CIKM), and IEEE International Conference on Data Mining (ICDM). His research interests include text mining, machine learning, ontology learning, and AI-based data analysis.

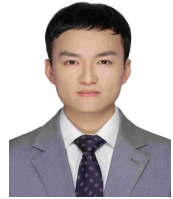

**Yinghui Huang.** He is an associate Professor of department of Management Science and Information Management, Wuhan University of Technology. His research interests encompass information systems, mental health, and health informatics. He has published research findings in prestigious international journals, including Information Processing & Management, with a particular focus on the machine behavior and psychology.